\documentclass[preprint]{ptephy_v1}

\preprintnumber{2203.15599} 
\usepackage{hyperref}

\usepackage{amsmath,amssymb,bm,float,mathrsfs}
\usepackage{graphicx}
\usepackage{dcolumn}
\usepackage{color}
\usepackage{hyperref}
\usepackage{comment}

  \newcommand{\al}[1]{\begin{align} #1 \end{align}}
  \def\dd{\mathrm{d}}
  
  \def\pd{\partial}
  \newcommand{\ave}[1]{\left\langle #1 \right\rangle}

\begin{document}

\title{Signature of primordial non-Gaussianity on 21-cm power spectrum from dark ages}


\author{Daisuke Yamauchi}
\affil{Faculty of Engineering, Kanagawa University, Kanagawa, 221-8686, Japan \email{yamauchi@jindai.jp}}

\begin{abstract}%
We study the signature of primordial non-Gaussianity imprinted on the power spectrum
of the 21-cm line differential brightness temperature during dark ages.
Employing the perturbative treatment of gravitational clustering, we quantitatively estimate 
the effects of the non-Gaussian and one-loop corrections on the 21-cm power spectrum.
The potential impact of the use of the 21-cm power spectrum for the constraint on
local-type primordial non-Gaussianity is investigated based on the Fisher matrix analysis.
Our results show that the 21-cm power spectrum for an array with 
a baseline of several tens of kilometers can constrain the primordial non-Gaussianity
to a level severer than that from cosmic microwave background measurements 
and its constraining power is stronger than that of the 21-cm bispectrum, while
in the ultimate situation the 21-cm bispectrum eventually becomes more powerful.
\end{abstract}

\subjectindex{xxxx, xxx}

\maketitle

\section{Introduction}

Inflation, the accelerated expansion phase of the early universe, has been widely studied as a standard paradigm 
that can naturally address the shortcomings of Big Bang cosmology. In particular, recent observations of anisotropies 
of cosmic microwave background (CMB) radiation, cosmic large-scale structure and so on strongly support the inflationary 
mechanism as the origin of primordial density fluctuations. 
Currently, these observations are consistent with a purely Gaussian distribution of fluctuations, but
possible small deviations from such a Gaussian primordial initial condition, called primordial non-Gaussianity, 
can be used to further constrain a bunch of inflation models and have extensively investigated.
The best constraint on primordial non-Gaussianity parametrized by the constant parameter $f_{\rm NL}^{\rm local}$~\cite{Komatsu:2001rj} 
are obtained from the CMB studies and are consistent with zero
(see \cite{Akrami:2019izv} for a recent constraint from Planck). However, the current CMB measurements are already 
reaching to the precision of the cosmic-variance limited one.

In this situation, high-precision deep-Universe exploration by large radio telescopes is of great importance.
Observations of the redshifted 21-cm line of neutral hydrogen (HI) open up a new window for observational cosmology 
(see \cite{Furlanetto:2006jb,Pritchard:2011xb} for a review).
One of the complementary ways to access primordial non-Gaussianity is to measure the spatial clustering behavior
of biased objects such as galaxies on large scales. This is because (local-type) primordial non-Gaussianity leads to the scale-dependent enhancement
of the large-scale clustering of biased objects due to the nonlinear interactions~\cite{Dalal:2007cu,Desjacques:2008vf}.
Next-generation radio galaxy surveys such as the Square Kilometre Array (SKA)~\footnote{The prospects of SKA to probe 
various aspects of cosmology have been summarized in \cite{SKA:2018ckk,SKA-JapanConsortiumCosmologyScienceWorkingGroup:2016hhj,Minoda:2022nso}.}
with the frequency range 50\,MHz--15.3\,GHz can reach the error of $f_{\rm NL}^{\rm local}$ close to unity
\cite{Camera:2013kpa,Raccanelli:2014kga,Yamauchi:2014ioa,Camera:2014bwa,Ferramacho:2014pua} (see also 
\cite{Yamauchi:2015mja,Yamauchi:2017ibz,Yamauchi:2021nsf} for constraining other types of primordial non-Gaussianities).
This is an important threshold to distinguish between single-field and multi-field inflation models.
Moreover, it has been shown that the simplest inflation model, namely single-field slow-roll inflation, generates
the considerably small amount of primordial non-Gaussianity $f_{\rm NL}^{\rm local}={\cal O}(0.01)$~\cite{Maldacena:2002vr}.
Therefore, other data set is needed to reach $f_{\rm NL}^{\rm local}<1$ frontier.

The last probe of the primordial non-Gaussianity is to use radio observations at less than $50\,{\rm MHz}$.
Such low-frequency observations allow us to map out the distribution of HI in the very deep Universe with the redshift ranges $30\leq z\leq 100$.
Since during these eras most scales remain linear and the nonlinear growth of structure 
is less effective in comparison with that in later epochs, we can in principle easily obtain predictable signals 
and a large number of Fourier samples which would drastically reduce the sample noises. 
In addition to these, we conduct a purely cosmological analysis avoiding astrophysical uncertainties because no stars are expected to 
form during dark ages.
Unfortunately, the radio signals at less than $10~{\rm MHz}$ cannot be measured from Earth due to the reflection of Earth's ionosphere. 
Therefore, there are many projects to measure the 21-cm line at dark ages from the far-side of the Moon, 
such as Dark Ages Polarimeter PathfindER 
(DAPPER, \cite{Burns:2021ndk}), Farside Array for Radio Science Investigations of the Dark ages and Exoplanets 
(FARSIDE, \cite{Burns:2021pkx}), Netherlands-China Low frequency Explorer (NCLE, \cite{2020AdSpR..65..856B}), 
and Lunar Crater Radio Telescope on the Far-Side of the Moon (LCRT, \cite{LCRT}).
Such moon-based instruments can avoid not only the ionospheric effects but also radio frequency interference which is
one of the severe systematics of the 21-cm observation.
Since the 21-cm fluctuations can extend to very small scales, the 21-cm measurements can provide us the information of 
not only the primordial non-Gaussianity but also the small-scale quantities of the primordial power spectrum such as
the running of the spectral index and other cosmological
observables~\cite{Loeb:2003ya,Lewis:2007kz,Mao:2008ug,Chen:2016zuu,Munoz:2016owz,Shiraishi:2016omb,Munoz:2018jwq}.

In this paper, we will study the fluctuations of the 21-cm line differential brightness temperature 
during dark ages as the ultimate probe of primordial inflation, in particular primordial non-Gaussianity.
Several studies have addressed this issue~\cite{Cooray:2006km,Pillepich:2006fj,Munoz:2015eqa,Meerburg:2016zdz,Xu:2016kwz,Floss:2022grj}. 
The authors in these literatures focused on the 21-cm bispectrum to constrain primordial non-Gaussianity
and pointed out that the secondary 21-cm bispectrum should be properly taken into account when their forecasts.
While the baryon and photon fluctuations are highly linear at the epoch of last-scattering, the perturbations in
the cold dark matter (CDM) and baryon fluids can significantly grow by $z\sim 100$.
They remain still small enough such that no bound structure such as stars has formed, but
gravitational nonlinear growth of structure gives non-negligible contributions to the density and velocity field.
Hence, the observed 21-cm brightness temperature depends on the baryon and velocity fluctuations nonlinearly.
In the present paper we study the effect of the nonlinear growth of structure and 
the signature of primordial non-Gaussianity imprinted on the 21-cm fluctuations,
especially focusing on the 21-cm power spectrum.
Naively thinking, the signature of primordial non-Gaussianity basically appears in the statistical properties of higher-order quantities,
and the power spectrum as a second-order statistics remains unchanged even if large primordial non-Gaussianity is presented.
However, as gravitational clustering develops, the coupling between Fourier modes of fluctuations becomes important and
the scale-dependent nonlinear growth appears due to the mode-coupling.
Since non-Gaussian density field intrinsically possesses a non-trivial mode-coupling, it can affect the late-time
evolution of the 21-cm power spectrum in the weakly nonlinear regime.

This paper is organized as follows.
In section \ref{sec:21-cm line during dark ages}, we first briefly review the 21-cm line signals during dark ages.
Expanding the 21-cm line differential brightness temperature up to the third order in the perturbative expansion, we then derive 
the statistical quantities such as the 21-cm bispectrum and 21-cm one-loop power spectrum, which contain
the effect of primordial non-Gaussianity.
In section \ref{sec:Fisher analysis}, we present Fisher forecasts on the amount of information available to constrain
primordial non-Gaussianity by combing the 21-cm power spectrum and the 21-cm bispectrum.
Finally, section \ref{sec:Summary} is devoted to the summary and discussion.

Throughout this paper, as our fiducial model we assume a $\Lambda$CDM cosmological mode with parameters:
$\Omega_{{\rm m},0}=0.3111$, $\Omega_{{\rm b},0}=0.0490$, $\Omega_{\Lambda}=0.6889$, $w=-1$, $h=0.6766$, $A_{\rm s}=2.105\times 10^{-9}$, $n_{\rm s}=0.9665$,
$k_{\rm pivot}=0.05$, and $Y_p=0.24$.

\section{21-cm line during dark ages}
\label{sec:21-cm line during dark ages}

In this section, we briefly review the 21-cm line signals during dark ages. 
We consider the spatial fluctuations of 21-cm line differential brightness temperature
in terms of the density and velocity fields up to the third order in perturbative expansion.
Based on the resultant expressions, we discuss the statistical quantities of
the 21-cm fluctuations such as bispectrum and power spectrum.
In particular, we will derive the 21-cm one-loop power spectrum including
the effect of the primordial bispectrum.
We then qualitatively estimate the impact of primordial non-Gaussianity 
and the use of the 21-cm power spectrum to probe primordial non-Gaussianity.

\subsection{21-cm brightness temperature}

The optical depth of the 21-cm line, $\tau$, is the function of the neutral hydrogen density $n_{\rm HI}=n_{\rm H}(1-x_{\rm e})$ and 
the gradient of the peculiar velocity field along the direction of propagation $\pd v_r/\pd r_{\rm phys}$ as
\al{
	\tau =&\frac{3c^3\hbar A_{10}n_{\rm HI}}{16k_{\rm B}\nu_{21}^2T_{\rm s}(\pd v_r/\pd r_{\rm phys})}
	\,,
}
where $A_{10}$ is the Einstein-A coefficient of the hyperfine transition, $\nu_{21}$ is the frequency corresponding to 21-cm line.
The brightness temperature observed today is given by
\al{
	T_{\rm b}:=T_{\rm CMB}e^{-\tau}+T_{\rm s}\left( 1-e^{-\tau}\right)
	\,.
}
Here $T_{\rm s}$ denotes the spin temperature, which can be evaluated during dark ages as
\al{
	T_{\rm s}=\frac{T_{\rm CMB}+y_{\rm c}T_{\rm gas}}{1+y_{\rm c}}
	\,,
}
with $y_{\rm c}=C_{10}T_\ast /A_{10}T_{\rm gas}$.
In this paper, the coefficient $C_{10}$ takes the form $C_{10}=n_{\rm H}\kappa_{10}^{\rm HH}(T_{\rm gas})$, 
in which we will use the fitting function of the matter temperature proposed in \cite{Kuhlen:2005cm}.
In the region of interest, the 21-cm transition is optically thin, hence the differential brightness temperature can be well approximated as
\al{
	T_{21}:=\frac{T_{\rm b}-T_{\rm CMB}}{1+z}\approx\frac{T_{\rm s}-T_{\rm CMB}}{1+z}\tau
	\,.
}
To evaluate the fluctuation of the 21-cm differential brightness temperature defined above, we first define 
the velocity perturbation $\delta_v ({\bm x},z):=-\pd_r v_r({\bm x},z)/{\cal H}(z)$ and the fluctuations of the matter temperature 
$\delta_{\rm T}({\bm x},z):=T_{\rm gas}({\bm x},z)/\overline T_{\rm gas}(z)-1$.
Throughout this paper, quantities with overline such as $\overline T_{\rm gas}$ and $\overline T_{\rm s}$ represent spatially averaged values
and should be evaluated through the background evolution.
With these, the observed brightness temperature can be expanded up to the third-order in the fluctuations as~\cite{Ali-Haimoud:2013hpa,Munoz:2015eqa}
\al{
	T_{21}=&\overline T_{21}\left( 1+\delta_v+\delta_v^2+\delta_v^3\right)
		+\left({\cal T}_{\rm b}\delta_{\rm b}+{\cal T}_{\rm T}\delta_{\rm T}\right)\left( 1+\delta_v+\delta_v^2\right)
	\notag\\
			&+\left({\cal T}_{\rm bb}\delta_{\rm b}^2+{\cal T}_{\rm bT}\delta_{\rm b}\delta_{\rm T}+{\cal T}_{\rm TT}\delta_{\rm T}^2\right)\left( 1+\delta_v\right)
	\notag\\
		&+{\cal T}_{\rm bbb}\delta_{\rm b}^3+{\cal T}_{\rm bbT}\delta_{\rm b}^2\delta_{\rm T}+{\cal T}_{\rm bTT}\delta_{\rm b}\delta_{\rm T}^2+{\cal T}_{\rm TTT}\delta_{\rm T}^3
	\,,\label{eq:T21 exp}
}
where $\overline T_{21}(z)$ and ${\cal T}_i(z)$ can be evaluated on the background and
the explicit expressions of these coefficients are given in Appendix~\ref{sec:Coefficients}.
In this expression, we have neglected the contributions from the fluctuations of the ionization fraction to the 21-cm fluctuation,
since the contribution of the ionization fraction is always suppressed by $\overline x_{\rm e}$.
We also have assumed the fluctuation of the hydrogen can be given by the baryon fluctuation, that is, $\delta n_{\rm H}/\overline n_{\rm H}\approx\delta_{\rm b}$.

As for the matter temperature, it has been shown in Ref.~\cite{Floss:2022grj} that when solving the evolution of the matter temperature
we need to take into account the fluctuation of the ionization fraction to reduce the error of the matter temperature fluctuation.
In this paper, we follow this strategy and require the additional equation describing the evolution of the ionization fraction.
Neglecting fluctuations of the CMB temperature,
the evolution equation for the fluctuation of the matter temperature is given by~\cite{Pillepich:2006fj}
\al{
	\dot\delta_{\rm T}-\frac{2}{3}\dot\delta_{\rm b}\frac{1+\delta_{\rm T}}{1+\delta_{\rm b}}
		=\Gamma_{\rm C}
			\biggl[
				\left(\frac{\overline T_{\rm CMB}}{\overline T_{\rm gas}}-1\right)\delta_x
				-\frac{\overline T_{\rm CMB}}{\overline T_{\rm gas}}\delta_{\rm T}
				-\delta_x\delta_{\rm T}
			\biggr]
	\,.\label{eq:deltaT eq}
}
where we have defined the Compton interaction rate as
\al{
	\Gamma_{\rm C}:=\frac{8\sigma_{\rm T}a_{\rm r}T_{\rm CMB}^4}{3m_{\rm e}}\frac{\overline x_{\rm e}}{1+\overline x_{\rm He}+\overline x_{\rm e}}
	\,,
}
with $\sigma_{\rm T}$, $a_{\rm r}$, and $m_{\rm e}$ being the Thomson cross-section, the radiation constant, and the election mass, respectively.
As for the ionization fraction, the evolution equation during the redshift range we consider in this paper can be written in the simple form~\cite{Ali-Haimoud:2013hpa}:
\al{
	\dot x_{\rm e}=-\alpha_{\rm B}(T_{\rm gas})n_{\rm H}x_{\rm e}^2
	\,.\label{eq:x_e eq}
}
Here the recombination coefficient $\alpha_{\rm B}$ is taken to be a fitting function of the matter temperature proposed in \cite{Seager:1999km}.
As mentioned before, we will solve Eqs.~\eqref{eq:deltaT eq} and \eqref{eq:x_e eq} simultaneously to obtain the precise time-evolution 
of the matter temperature.
The perturbed equations for these equations to solve are shown in Appendix \ref{sec:Evolution equations}.

Even if the primordial fluctuation is well described by linear theory, the nonlinearity of the gravitational dynamics eventually dominates
and we must correctly take into account the nonlinear growth of fluctuations.
For the scales of interest, the nonlinear evolution is rather moderate and perturbative treatment is still valid.
Hence, we decompose the perturbations into a piece linear in the initial condition, $\delta_{\rm X}^{(1)}$, and higher-order pieces,
$\delta_{\rm X}^{(n)}$, resulting
from nonlinear gravitational evolution, namely $\delta_{\rm X}=\delta^{(1)}_{\rm X}+\delta^{(2)}_{\rm X}+\cdots$.
As for the baryon fluctuation $\delta_{\rm b}$, the $n$-th order solution can be formally described by the convolution of the first-order one as
\al{
	\delta_{\rm b}^{(n)} ({\bm k},z)
		=&\int\frac{\dd^3{\bm p}_1\cdots\dd^n{\bm p}_2}{(2\pi )^{n-1}}\delta_{\rm D}^3({\bm k}-{\bm p}_1+\cdots -{\bm p}_n)
	\notag\\
	&\times
				F_n^{({\rm sym})}({\bm p}_1,\cdots,{\bm p}_n)\delta_{\rm b}^{(1)}({\bm p}_1,z)\cdots\delta_{\rm b}^{(1)}({\bm p}_n,z)
	\,,
}
where $F_n^{({\rm sym})}$ denotes the symmetrized $n$-th order perturbative kernels, which can be written in the matter-dominated Universe.
The second- and third-order kernels of $F_n^{({\rm sym})}$ are shown in Eqs.~\eqref{eq:F2} and \eqref{eq:F3}.
Although the baryon perturbative kernels are generally different from the CDM ones because of the baryon sound speed,
in this paper we simply use the CDM perturbative kernels as the baryon perturbative kernels, because the main purpose of
this paper is to show the impact of the higher-order perturbations on the constraint of the primordial non-Gaussianity by using
not only the 21-cm bispectrum but also the one-loop 21-cm power spectrum.
In addition, the velocity perturbation $\delta_v$ can be written in the similar form.
The Fourier transform of $\delta_v$ is written in terms of the Fourier component of the velocity divergence 
$\theta_{\rm b}:=\nabla\cdot{\bm v}_{\rm b}$ as
\al{
	\delta_v ({\bm k},z)=-\mu^2\frac{\theta_{\rm b}({\bm k},z)}{{\cal H}(z)}
	\,,
} 
with $\mu={\bm k}\cdot\widehat{\bm n}/k$.
The velocity divergence can be expanded in terms of the kernel functions in the same manner as the baryon fluctuation and the $n$-th order solution
is written as
\al{
	\theta_{\rm b}^{(n)} ({\bm k},z)
		=&{\cal H}(z)f(z)\int\frac{\dd^3{\bm p}_1\cdots\dd^3{\bm p}_2}{(2\pi )^{n-1}}\delta_{\rm D}^3({\bm k}-{\bm p}_1-\cdots-{\bm p}_n)
	\notag\\
	&\times
				G_n^{({\rm sym})}({\bm p}_1,\cdots,{\bm p}_n)\delta_{\rm b}^{(1)}({\bm p}_1,z)\cdots\delta_{\rm b}^{(1)}({\bm p}_n,z)
	\,.
}
Here $G_n^{({\rm sym})}$ is the symmetrized $n$-th order perturbative kernels of the velocity divergence and we have introduced the growth rate 
$f:=\dd\ln\delta^{(1)}/\dd\ln a$.
The second- and third-order kernel of $G_n^{({\rm sym})}$ are shown in Eqs.~\eqref{eq:G2} and \eqref{eq:G3}.
In particular, at the first order the continuity equation leads to $\delta_v^{(1)}=f\mu^2\delta_{\rm b}^{(1)}$.

In order to investigate the 21-cm fluctuation, we need to solve the perturbations of the matter temperature and ionization fraction simultaneously.
We assume that these fluctuations can be well approximated by the baryon fluctuation in the form:~\cite{Pillepich:2006fj,Munoz:2015eqa,Floss:2022grj}
\al{
	&\delta_{\rm X}^{(1)}({\bm x},z)
		=C_{{\rm X},1}(z)\delta_{\rm b}^{(1)}({\bm x},z)
	\,,\label{eq:delta_X1}\\
	&\delta_{\rm X}^{(2)}({\bm x},z)
		=C_{{\rm X},2}^{(1)}(z)\Bigl\{[\delta_{\rm b}^{(1)}({\bm x},z)]^2-\ave{[\delta_{\rm b}^{(1)}({\bm x},z)]^2}\Bigr\}+C_{{\rm X},2}^{(2)}(z)\delta_{\rm b}^{(2)}({\bm x},z)
	\,,\label{eq:delta_X2}\\
	&\delta_{\rm X}^{(3)}({\bm x},z)
		=C_{{\rm X},3}^{(1)}(z)\Bigl\{[\delta_{\rm b}^{(1)}({\bm x},z)]^3-\ave{[\delta_{\rm b}^{(1)}({\bm x},z)]^3}\Bigr\}
	\notag\\
	&\qquad\qquad\qquad
			+C_{{\rm X},3}^{(2)}\Bigl\{\delta_{\rm b}^{(1)}({\bm x},z)\delta_{\rm b}^{(2)}({\bm x},z)-\ave{\delta_{\rm b}^{(1)}({\bm x},z)\delta_{\rm b}^{(2)}({\bm x},z)}\Bigr\}
			+C_{{\rm X},3}^{(3)}\delta_{\rm b}^{(3)}({\bm x},z)
	\,,\label{eq:delta_X3}
}
where ${\rm X}={\rm T}$ and $x$.
This assumption is valid as long as $\delta_{\rm b}^{(n)}$ grows independently of the position ${\bm x}$.
Indeed, for era of our interests, $z\sim 30$--$150$, we expect that the large-scale fluctuation of baryons behaves like CDM and
the large-scale baryon fluctuation can be well approximated by the scale-independent growth, though the deviation due to the pressure appears on small scales.
With these assumptions, the fluctuations of the 21-cm differential brightness temperature can be described in terms of the baryon fluctuation and
the velocity perturbation as~\cite{Munoz:2015eqa,Floss:2022grj}
\al{
	\delta T_{21}=&\alpha_1 \delta_{\rm b}^{(1)}+\overline T_{21}\delta_v^{(1)}
	\notag\\
		&+\alpha_2^{(2)} \delta_{\rm b}^{(2)}
			+\alpha_2^{(1)} [\delta_{\rm b}^{(1)}]^2
			+\alpha_1 \delta_{\rm b}^{(1)}\delta_v^{(1)}
			+\overline T_{21}\Bigl\{ \delta_v^{(2)}+[\delta_v^{(1)}]^2\Bigr\}
	\notag\\
		&+\alpha_3^{(3)}\delta_{\rm b}^{(3)}
			+\alpha_3^{(2)}\delta_{\rm b}^{(1)}\delta_{\rm b}^{(2)}
			+\alpha_3^{(1)} [\delta_{\rm b}^{(1)}]^3
			+\alpha_2^{(2)}\delta_{\rm b}^{(2)}\delta_v^{(1)}
			+\alpha_1\delta_{\rm b}^{(1)}\delta_v^{(2)}
	\notag\\
		&+\overline T_{21}\Big\{
				\delta_v^{(3)}
				+2\delta_v^{(1)}\delta_v^{(2)}
				+[\delta_v^{(1)}]^3
			\Bigr\}
	\,.\label{eq:delta T21}
}
Here the time-dependent functions $\alpha_1$ and $\alpha_n^{(m)}$ are rewritten in terms of the coefficients in Eqs.~\eqref{eq:T21 exp}--\eqref{eq:delta_X3} 
as
\al{
	&\alpha_1 ={\cal T}_{\rm b}+C_{{\rm T},1}{\cal T}_{\rm T}
	\,,\\
	&\alpha_2^{(1)} ={\cal T}_{\rm bb}+C_{{\rm T},1}{\cal T}_{\rm bT}+C_{{\rm T},2}^{(1)}{\cal T}_{\rm T}+[C_{{\rm T},1}]^2{\cal T}_{\rm TT}
	\,,\\
	&\alpha_2^{(2)} ={\cal T}_{\rm b}+C_{{\rm T},2}^{(2)}{\cal T}_{\rm T}
	\,,
}
and 
\al{
	&\alpha_3^{(1)}={\cal T}_{\rm bbb}+C_{{\rm T},1}{\cal T}_{\rm bbT}+C_{{\rm T},2}^{(1)}{\cal T}_{\rm bT}+[C_{{\rm T},1}]^2{\cal T}_{\rm bTT}
	\notag\\
	&\qquad\qquad\qquad
		+C_{{\rm T},3}^{(1)}{\cal T}_{\rm T}+2C_{{\rm T},1}C_{{\rm T},2}^{(1)}{\cal T}_{\rm TT}+[C_{{\rm T},1}]^3{\cal T}_{\rm TTT}
	\,,\\
	&\alpha_3^{(2)}=2{\cal T}_{\rm bb}+\left( C_{{\rm T},1}+C_{{\rm T},2}^{(2)}\right){\cal T}_{\rm bT}+C_{{\rm T},3}^{(2)}{\cal T}_{\rm T}+2C_{{\rm T},1}C_{{\rm T},2}^{(2)}{\cal T}_{\rm TT}
	\,,\\
	&\alpha_3^{(3)}={\cal T}_{\rm b}+C_{{\rm T},3}^{(3)}{\cal T}_{\rm T}
	\,.
}
Based on these results, we obtain the Fourier component of the 21-cm fluctuation Eq.~\eqref{eq:delta T21} up to the third order as 
\al{
	&\delta T_{21}({\bm k},z)=Z_1({\bm k},z)\delta_{\rm b}({\bm k},z)
	\notag\\
	&\qquad
		+\int\frac{\dd^3{\bm p}_1\dd^3{\bm p}_2}{(2\pi)^3}\delta_{\rm D}^3({\bm k}-{\bm p}_1-{\bm p}_2)
			Z_2({\bm p}_1,{\bm p}_2;z)\delta_{\rm b}({\bm p}_1,z)\delta_{\rm b}({\bm p}_2,z)
	\notag\\
	&\qquad
		+\int\frac{\dd^3{\bm p}_1\dd^3{\bm p}_2\dd^3{\bm p}_3}{(2\pi)^6}\delta_{\rm D}^3({\bm k}-{\bm p}_1-{\bm p}_2-{\bm p}_3)
			Z_3({\bm p}_1,{\bm p}_2,{\bm p}_3;z)\delta_{\rm b}({\bm p}_1,z)\delta_{\rm b}({\bm p}_2,z)\delta_{\rm b}({\bm p}_3,z)
	\notag\\
	&\qquad
		+\cdots
	\,,\label{eq:Fourier delta T21}
}
where $Z_1$, $Z_2$, and $Z_3$ denote the linear-, second-, and third-order perturbative kernels, which is written as
(see \cite{Scoccimarro:1999ed} for a galaxy fluctuation)
\al{
	&Z_1({\bm k})=\alpha_1 +\overline T_{21}f\mu^2
	\,,\\
	&Z_2({\bm k}_1,{\bm k}_2)
		=\alpha_2^{(2)} F_2^{({\rm sym})}({\bm k}_1,{\bm k}_2)-\overline T_{21}f\mu^2 G_2^{({\rm sym})}({\bm k}_1,{\bm k}_2)
	\notag\\
	&\qquad\qquad\qquad\qquad
			+\frac{1}{2}f\alpha_1 \left(\mu_1^2+\mu_2^2\right) +\overline T_{21}f^2\mu_1^2\mu_2^2 +\alpha_2^{(1)}
	\,,\label{eq:Z2}\\
	&Z_3({\bm k}_1,{\bm k}_2,{\bm k}_3)
		=\alpha_3^{(3)} F_3^{({\rm sym})}({\bm k}_1,{\bm k}_2,{\bm k}_3)-\overline T_{21}f\mu^2 G_3^{({\rm sym})}({\bm k}_1,{\bm k}_2,{\bm k}_3)
	\notag\\
	&\qquad\qquad\qquad\qquad
		+\frac{1}{3}\Bigl[\left(\alpha_3^{(2)}+\alpha_2^{(2)}f\mu_1^2\right) F_2^{({\rm sym})}({\bm k}_2,{\bm k}_3)+(\text{perms})\Bigr]
		+\alpha_3^{(1)}
	\notag\\
	&\qquad\qquad\qquad\qquad
		-\frac{f}{3}\Bigl[\left(\alpha_1+2\overline T_{21}f\mu_1^2\right) \mu_{23}^2G_2^{({\rm sym})}({\bm k}_2,{\bm k}_3)+(\text{perms})\Bigr]
		+\overline T_{21}f^3\mu_1^2\mu_2^2\mu_3^2
	\,.\label{eq:Z3}
}
where $\mu:={\bm k}\cdot\widehat{\bm n}/k$, $\mu_i ={\bm k}_i\cdot\widehat{\bm n}/k_i$, and 
$\mu_{ij}={\bm k}_{ij}\cdot\widehat{\bm n}/k_{ij}$ with ${\bm k}_{ij}={\bm k}_i+{\bm k}_j$.
We further assume that the time-evolution of the baryon fluctuation is exactly same as the CDM one during the matter dominated era, 
namely $\delta_{\rm b}^{(n)}\propto a^n$.
Under this assumption, the growth rate $f$ becomes unity.

\subsection{Primordial fluctuations}

Once the statistical nature of primordial curvature perturbation $\zeta$ is specified, the linear density field is determined through
\al{
	&\delta_{\rm b}^{(1)}({\bm k},z)
		={\cal M}(k,z)\zeta ({\bm k})
	\,,
}
where the function ${\cal M}(k,z)$ is defined as ${\cal M}(k,z)=2k^2T(k)D(z)/5H_0^2\Omega_{\rm m,0}$
with $D(z)$ and $T(k)$ being the linear growth rate and matter transfer function, respectively.
In the present analysis we work in terms of primordial curvature perturbation $\zeta$ with power spectrum
\al{
	\ave{\zeta ({\bm k}_1)\zeta ({\bm k}_2)}=(2\pi)^3\delta_{\rm D}^3({\bm k}_{12})P_\zeta (k)
	\,,
}
bispectrum, and trispectrum
\al{
	&\ave{\zeta ({\bm k}_1)\zeta ({\bm k}_2)\zeta ({\bm k}_3)}=(2\pi)^3\delta_{\rm D}^3({\bm k}_{123})B_\zeta (k_1,k_2,k_3)
	\,,\label{eq: prim bi}\\
	&\ave{\zeta ({\bm k}_1)\zeta ({\bm k}_2)\zeta ({\bm k}_3)\zeta ({\bm k}_4)}=(2\pi)^3\delta_{\rm D}^3({\bm k}_{1234})T_\zeta (k_1,k_2,k_3,k_4)
	\,,
}
with ${\bm k}_{ij}={\bm k}_i+{\bm k}_j$, ${\bm k}_{ijk}={\bm k}_i+{\bm k}_j+{\bm k}_k$, and ${\bm k}_{ijk\ell}={\bm k}_i+{\bm k}_j+{\bm k}_k+{\bm k}_\ell$.

Since primordial non-Gaussianity reflects the fundamental interactions and nonlinear processes involved
during and after inflation, it can bring the insights into the generating mechanism of primordial fluctuations.
In the simplest case, the curvature perturbation $\zeta$ can be expanded in terms of the purely Gaussian
variable $\zeta_{\rm g}$ as~\cite{Komatsu:2001rj}
\al{
	\zeta =\zeta_{\rm g}+\frac{3}{5}f^{\rm local}_{\rm NL}\zeta_{\rm g}^2+\cdots
	\,,
}
which leads to the primordial bispectrum of the form:
\al{
	B_\zeta (k_1,k_2,k_3)=\frac{6}{5}f_{\rm NL}^{\rm local}\biggl[ P_\zeta (k_1)P_\zeta (k_2)+(\text{2\ perms})\biggr]
	\,.\label{eq:local fNL def}
}
Here $f_{\rm NL}^{\rm local}$ is called the nonlinearity parameter for the local-type and usually assumed to be
constant (see \cite{Yamauchi:2021nsf} for the scale-dependence of the local-form of the nonlinear parameters).
Although the generalization to other types of primordial non-Gaussianity such as equilateral- and orthogonal-types 
is straightforward, in this paper we focus on the local-type primordial non-Gaussianity Eq.~\eqref{eq:local fNL def}.
This is mainly because other types of primordial non-Gaussianities do not induce the strong scale-dependence
in the 21-cm power spectrum because of weaker mode-correlations between small and large Fourier modes,
as shown in the subsequent analysis.

\subsection{Bispectrum of 21-cm fluctuations}

The lowest order contribution from primordial non-Gaussianity to the bispectrum of the fluctuation of the 21-cm differential brightness temperature
can be obtained by considering the linear term in Eq.~\eqref{eq:Fourier delta T21}.
We define the bispectrum of the 21-cm fluctuations as
\al{
	\ave{\delta T_{21}({\bm k}_1)\delta T_{21}({\bm k}_2)\delta T_{21}({\bm k}_3)}
		=(2\pi)^3\delta_{\rm D}^3({\bm k}_1+{\bm k}_2+{\bm k}_3)B({\bm k}_1,{\bm k}_2,{\bm k}_3)
	\,.
}
Given the primordial bispectrum $B_\zeta (k_1,k_2,k_3)$ which is defined in Eq.~\eqref{eq: prim bi},
one can find the resulting tree-level 21-cm bispectrum due to the primordial bispectrum as
\al{
	&B^{\rm prim}({\bm k}_1,{\bm k}_2,{\bm k}_3)
		=B_\zeta (k_1,k_2,k_3)\prod_{i=1}^3Z_1({\bm k}_i){\cal M}_\zeta (k_i)
	\,.
}
Due to the nonlinear growth of the perturbations under gravity, the 21-cm fluctuations from dark ages
are not exactly linear and the non-negligible secondary non-Gaussian signals appear.
To extract the signals of primordial non-Gaussianity from the 21-cm bispectrum, we need to
accurately model the secondary contributions. 
In our notation, the secondary bispectrum can be written as
\al{
	&B^{\rm sec}({\bm k}_1,{\bm k}_2,{\bm k}_3)
		=2Z_1({\bm k}_1)Z_1({\bm k}_2)Z_2({\bm k}_1,{\bm k}_2)P_\delta ({\bm k}_1)P_\delta ({\bm k}_2)
		+(\text{2\ perms.})
	\,.
}
It was shown in Refs.~\cite{Pillepich:2006fj,Munoz:2015eqa} that the secondary contributions to the 21-cm bispectrum 
give the several order of magnitude larger than the primordial one and hence the secondary contributions in the 21-cm bispectrum 
dominate the signals. 

\subsection{One-loop power spectrum of 21-cm fluctuations and signature of primordial non-Gaussianity}
\label{sec:One-loop power spectrum of 21-cm fluctuations and signature of primordial non-Gaussianity}

In this section, by using the perturbative expansion of the 21-cm fluctuations \eqref{eq:delta T21}, we construct
the power spectrum up to the one-loop level. 
Defining the power spectrum of the 21-cm fluctuation through
\al{
	\ave{\delta T_{21}({\bm k})\delta T_{21}({\bm k}^\prime)}=(2\pi)^3\delta_{\rm D}({\bm k}+{\bm k}^\prime )P(k)
	\,,
}
we obtain the power spectrum of the 21-cm fluctuation up to the one-loop order can be written as \cite{Taruya:2008pg}
\al{
	P({\bm k};z)=P^{(11)}({\bm k};z)+P^{(12)}({\bm k};z)+\Bigl[P^{(22)}({\bm k};z)+P^{(13)}({\bm k};z)\Bigr]
	\,,
}
where
\al{
	&P^{(11)}({\bm k})
		=Z_1^2({\bm k})P_{\delta}({\bm k})
	\,,\\
	&P^{(12)}({\bm k})
		=2Z_1({\bm k})\int\frac{\dd^3{\bm p}}{(2\pi)^3}Z_2({\bm p},{\bm k}-{\bm p})B_\delta (k,p,|{\bm k}-{\bm p}|)
	\,,\\
	&P^{(22)}({\bm k})
		=2\int\frac{\dd^3{\bm p}}{(2\pi)^3}\bigl[Z_2({\bm p},{\bm k}-{\bm q})\bigr]^2P_\delta (p)P_\delta (|{\bm k}-{\bm p}|)
	\notag\\
	&\qquad\qquad\qquad
			+\int\frac{\dd^3{\bm p}\dd^3{\bm q}}{(2\pi)^6}Z_2({\bm p},{\bm k}-{\bm p})Z_2({\bm q},-{\bm k}-{\bm q};z)
				T_\delta ({\bm p},{\bm k}-{\bm p},{\bm q},-{\bm k}-{\bm q})
	\,,\label{eq:P22}\\
	&P^{(13)}({\bm k})
		=6Z_1({\bm k})\int\frac{\dd^3{\bm p}}{(2\pi)^3}Z_3({\bm k},{\bm p},-{\bm p})P_\delta (k)P_\delta (p)
	\notag\\
	&\qquad\qquad\qquad
			+2Z_1({\bm k})\int\frac{\dd^3{\bm p}\dd^3{\bm q}}{(2\pi)^6}Z_3({\bm p},{\bm q},{\bm k}-{\bm p}-{\bm q})
				T_\delta (-{\bm k},{\bm p},{\bm q},{\bm k}-{\bm p}-{\bm q})
	\,.\label{eq:P13}
}
Here $P_\delta (k)$, $B_\delta (k_1,k_2,k_3)$, and $T_\delta ({\bm k}_1,{\bm k}_2,{\bm k}_3,{\bm k}_4)$ denote 
the power-, bi-, and tri-spectrum for the baryon fluctuation, which are related to the primordial spectrum through
$P_\delta (k)={\cal M}(k)P_\zeta (k)$, $B_\delta (k_1,k_2,k_3)={\cal M}(k_1){\cal M}(k_2){\cal M}(k_3)B_\zeta (k_1,k_2,k_3)$, and
$T_\delta (k_1,k_2,k_3,k_4)={\cal M}(k_1){\cal M}(k_2){\cal M}(k_3){\cal M}(k_4)T_\zeta (k_1,k_2,k_3,k_4)$.
For simplicity, hereafter we neglect the contributions from the primordial trispectrum, but the generalization is straightforward.
We note that, in the high momentum limit of the loop integration, there exists the natural cutoff corresponding to 
the baryonic Jeans scale. Hence, we expect that the momentum integration in this limit always 
gives the finite result.

\begin{figure}
\includegraphics[width=150mm]{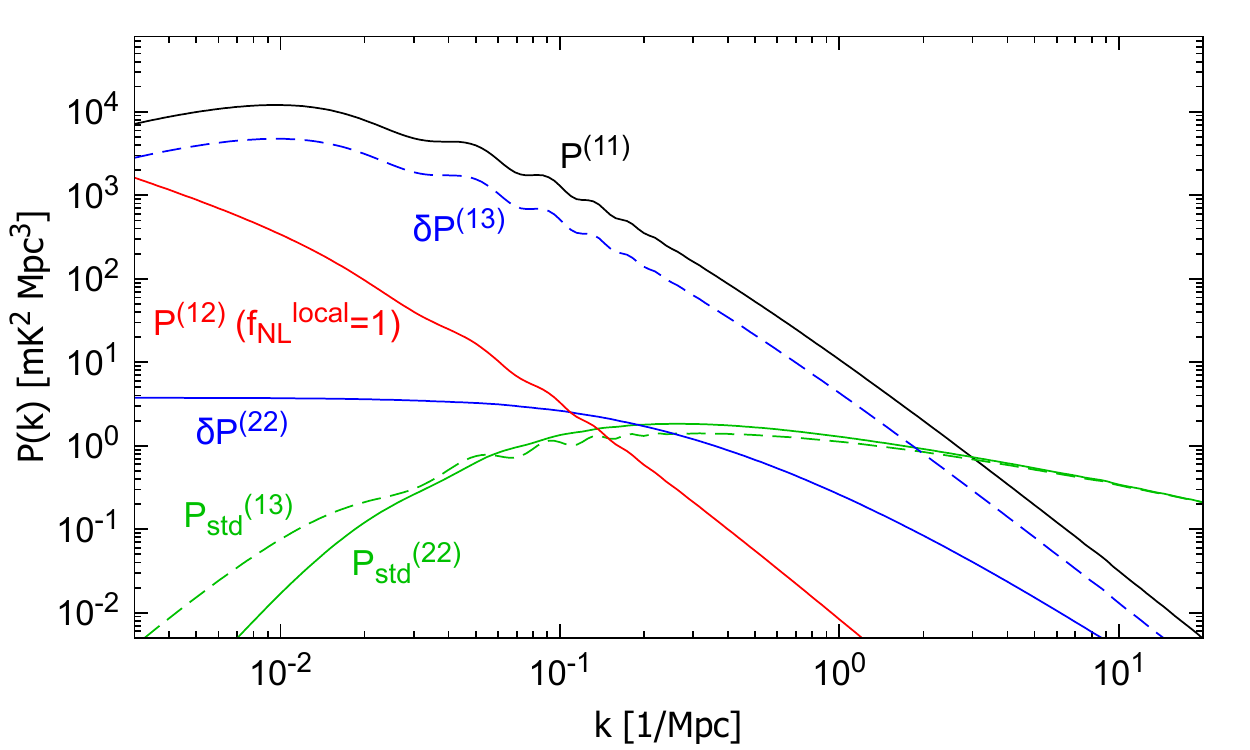}
\vspace{-2mm}
\caption{
The 21-cm power spectrum at $z=30$ with $\mu=0$.
The different colors represent the different contributions: linear term $P^{(11)}$ (black solid), standard one-loop term 
$P^{(22)}$ (green solid) and $P^{(13)}$  (green dashed), one-loop correction term $\delta P^{(22)}$ (blue solid) and $\delta P^{(13)}$ (blue dashed), 
and contribution from the primordial bispectrum $P^{(12)}$ (red solid) with
$f_{\rm NL}^{\rm local}=1$.
}
\label{fig:pow_z30_mu0}
\end{figure}

\begin{figure}
\includegraphics[width=210mm]{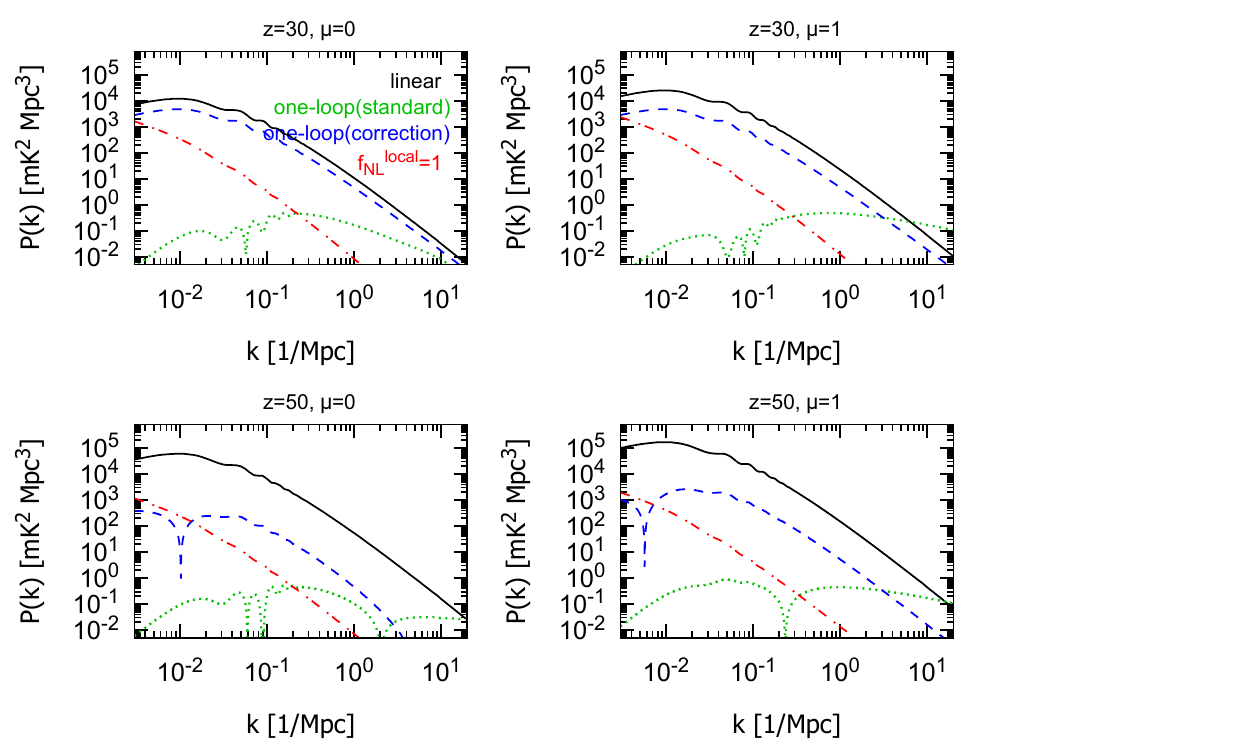}
\vspace{-5mm}
\caption{
The 21-cm power spectra for each contribution: linear term (black solid), standard one-loop term (green dotted), 
one-loop correction term (blue dashed), and bispectrum contribution (red dot-dashed).
}
\label{fig:pow}
\end{figure}

It is known that as for the standard matter perturbations, the one-loop matter power spectrum can be written in terms of only
$F_n^{({\rm sym})}$ and $G_n^{({\rm sym})}$.
In contrast, Eqs.~\eqref{eq:P22} and \eqref{eq:P13} with Eqs.~\eqref{eq:Z2} and \eqref{eq:Z3} imply that the 21-cm one-loop power 
spectra include the additional nontrivial contributions.
To estimate these effects separately, we decompose the one-loop terms $P^{(22)}$ and $P^{(13)}$
into two pieces:
\al{
	&P^{(22)}({\bm k})
		=P^{(22)}_{\rm std}({\bm k})+\delta P^{(22)}({\bm k})
	\,,\\
	&P^{(13)}({\bm k})
		=P^{(13)}_{\rm std}({\bm k})+\delta P^{(13)}({\bm k})
	\,.
}
Here, we refer to the terms related to only $F_n^{({\rm sym})}$ and $G_n^{({\rm sym})}$ as ``standard'' terms and others
as ``correction'' terms.
$P^{(22)}_{\rm std}$ and $P^{(13)}_{\rm std}$ denote the standard one-loop spectra, 
whose explicit forms are shown in Appendix \ref{sec:Standard one-loop power matter spectrum}.
The correction term of the one-loop power spectrum from the auto-correlation of the second-order perturbation is given by
\al{
	\delta P^{(22)}({\bm k})
		=&2\frac{k^3}{(2\pi)^2}\int_0^\infty\dd x x^2P_\delta (kx)\int_{-1}^1\dd\nu 
			P_\delta\bigl(k\sqrt{1+x^2-2\nu x}\,\bigr)
	\notag\\
	&\qquad\times
			\Sigma (x,\nu ;\mu)\Bigl[ 2\Xi (x,\nu ;\mu) +\Sigma (x,\nu ;\mu)\Bigr]
	\,,\label{eq:delta P22}
}
where
\al{
	&\Xi (x,\nu ;\mu)
		=\alpha_2^{(2)}\frac{3x+7\nu-10\nu^2x}{14x(1+x^2-2x\nu )}
				-\overline T_{21}\mu^2\frac{-x+7\nu -6\nu^2 x}{14x(1+x^2-2x\nu )}
	\,,\label{eq:Xi def}\\
	&\Sigma (x,\nu ;\mu)
		=\alpha_2^{(1)}+\frac{1}{2}\alpha_1\biggl[\frac{1}{2}(1-\mu^2)(1-\nu^2)+\mu^2\nu^2+\frac{x^2(1-\mu^2)(1-\nu^2)+2\mu^2(1-x\nu)^2}{2(1+x^2-2x\nu )}\biggr]
	\notag\\
	&\qquad
		+\overline T_{21}\frac{3x^2(1-\mu^2)^2(1-\nu^2)^2+4\mu^2(1-\mu^2)(1-\nu^2)(1-6x\nu +6x^2\nu^2)+8\mu^4\nu^2(1-x\nu)^2}{8(1+x^2-2x\nu )}
	\,,\label{eq:Sigma def}
}
Here, we have used the notation $x=p/k$ and $\nu=\widehat{\bm k}\cdot\widehat{\bm p}$.
On the other hand, the correction term from the cross-correlation between the linear- and third-order perturbations can be written as
\al{
	\delta P^{(13)}({\bm k})
		=&2Z_1({\bm k})\Upsilon_0\sigma_0^2P_\delta (k)
	\notag\\
		&-Z_1({\bm k})\frac{k^3}{(2\pi)^2}P_\delta (k)\int_0^\infty\dd x x^2P_\delta (kx)
				\Bigl[\alpha_1{\cal G}_1(x;\mu)+2\overline T_{21}\mu^2{\cal G}_2(x;\mu)\Bigr]
	\,,\label{eq:delta P13}
}
where $\sigma_0^2:=\int_0^\infty\dd pp^2P_\delta (p)/2\pi^2$,
\al{
	\Upsilon_0=\left( 3\alpha_3^{(1)}+\frac{34}{21}\alpha_3^{(2)}+\frac{18}{35}\alpha_2^{(2)}\right) +\left(\frac{8}{105}\alpha_2^{(2)}+\frac{3}{5}\overline T_{21}\right)\mu^2
	\,,
}
and ${\cal G}_1$ and ${\cal G}_2$ are defined as
\al{
	&{\cal G}_1(x;\mu)=\frac{1}{168}
			\biggl[
			 -\frac{18(1-3\mu^2)}{x^2}+66+634\mu^2-6x^2\left( 3x^2-11\right)\left( 1-3\mu^2\right)
	\notag\\
	&\qquad\qquad
			+\frac{9}{x^3}\left( x^2-1\right)^4\left( 1-3\mu^2\right)\ln\biggl|\frac{x+1}{x-1}\biggl|\,
			\biggr]
	\,,\\
	&{\cal G}_2(x;\mu)=\frac{1}{4480x^5}
			\biggl[
				6x\left(1+x^2\right)\left( 15-100x^2+298x^4-100x^6+15x^8\right)
	\notag\\
	&\qquad
				+4x\left( 15+35x^2+2334x^4-1410x^6+915x^8-225x^{10}\right)\mu^2
	\notag\\
	&\qquad
				+10x\left( x^2-1\right)^2\left( 9+15x^2-145x^4+105x^6\right)\mu^4
	\notag\\
	&\qquad
				+15\left( x^2-1\right)^4
					\Bigl\{
						3\left(x^2-1\right)^2+2\left(1+6x^2-15x^4\right)\mu^2
						+\left( 3+10x^3+35x^4\right)\mu^4
					\Bigr\}\ln\biggl|\frac{x+1}{x-1}\biggl|\,
			\biggr]
	\,.
}
The derivation of the last term in Eq.~\eqref{eq:delta P13} is shown in Appendix \ref{sec:Standard one-loop power matter spectrum}.
We show in Fig.~\ref{fig:pow_z30_mu0} the 21-cm power spectra at $z=30$ with $\mu=0$ for each contributions: $P^{(11)}$ (black solid), $P^{(22)}_{\rm std}$
(green solid), $P_{\rm std}^{(13)}$ (green dashed), $\delta P^{(22)}$ (blue solid), $\delta P^{(13)}$ (blue dashed), and $P^{(12)}$ (red solid) with $f_{\rm NL}^{\rm local}=1$.
To see the dependence on the redshift and the directional cosine, in Fig.~\ref{fig:pow} we show the 21-cm power spectra of linear term (black solid), 
standard one-loop term (green dotted), one-loop correction term (blue dashed), and bispectrum contribution (red dot-dashed),
for $z=30$ (upper panel) and $50$ (lower panel) with $\mu=0$ (left panel), $1$ (right panel).
\\

In order to study the one-loop contributions to the power spectrum for the 21-cm fluctuation, we would like to examine
their asymptotic behavior of the short and long wavelength limits in the loop integrals as done in \cite{Makino:1991rp}.
To do this, we split the one-loop contributions into two pieces: that from the momentum integration for $p\gg k$, namely $x\gg 1$,
(Hereafter we call it UV region.) and that from the integration for $p\ll k$, namely $x\ll 1$, (IR region), for fixed $k$, respectively. 
In particular, it was shown in \cite{Makino:1991rp} that in the standard perturbation theory the leading terms from 
the one-loop matter power spectrum in the IR limit are exactly canceled out (see also \cite{Hirano:2020dom}).
Hereafter, we will extend their analysis to the 21-cm one-loop power spectrum.

Let us first consider the long-wavelength (IR) contributions ($x\ll 1$) in the standard one-loop terms.
The sum of the standard parts, $P_{\rm std}^{(22)}+P_{\rm std}^{(13)}$, in this limit reduces to
\al{
	P^{(22)}_{\rm std}({\bm k})+P^{(13)}_{\rm std}({\bm k})
		\stackrel{\rm{IR}}{\approx}
				\frac{2}{3}\bigg[\left([\alpha_2^{(2)}]^2-\alpha_1\alpha_3^{(3)}\right) +\mu^2\left(\cdots \right)\biggr]
				\frac{k^2}{(2\pi)^2}P_\delta (k)\int_{p\ll k}\dd pP_\delta (p)
	\,.\label{eq:standard sum}
}
This result shows that the leading terms of the standard one-loop contributions are not exactly canceled out
but is only suppressed by the factor $([\alpha_2^{(2)}]^2-\alpha_1\alpha_3^{(3)})$.
Although this value is small but in general nonzero, in the low redshift regime ($z\lesssim 30$)
it can be shown to approach to zero because the matter temperature at low redshifts can be determined
by $T_{\rm gas}\propto n_{\rm H}^{3/2}$, which leads to $\alpha_1\approx\alpha_2^{(2)}\approx\alpha_3^{(3)}$.
Indeed, this behavior can be seen in Fig.~\ref{fig:pow_z30_mu0} and the upper left panel of Fig.~\ref{fig:pow}.
Eq.~\eqref{eq:standard sum} also shows that the sum of the standard terms is in proportion to $k^{2+n_{\rm IR}}$, which 
is different from the results shown in \cite{Makino:1991rp}, where we have introduced
the small-scale spectral tilt defined as $n_{\rm IR}=\dd\ln P_\delta/\dd\ln k$.
Hence, as seen in the lower panels of Fig.~\ref{fig:pow}, in high redshift regime ($z\gtrsim 50$) 
the standard one-loop contributions dominate the signal on scales longer than naively expected.
Let us move on the short-wavelength (UV) contributions ($x\gg 1$). In this limit, the sum of the standard terms reduces to
\al{
	P^{(22)}_{\rm std}({\bm k})+P^{(13)}_{\rm std}({\bm k})
		\stackrel{\rm{UV}}{\approx}
				Z_1({\bm k})\left( -\frac{122}{315}\alpha_3^{(3)}+\frac{6}{5}\overline T_{21}\mu^2\right)
				\frac{k^2}{(2\pi)^2}P_\delta (k)\int_{p\gg k}\dd pP_\delta (p)
	\,,
}
implying that the UV contribution of the standard parts
is suppressed by $k^2$ as usual.
\\

Next, we study the asymptotic behaviors of the correction terms of the one-loop contributions, $\delta P^{(22)}$ and $\delta P^{(13)}$, 
defined in Eqs.~\eqref{eq:delta P22} and \eqref{eq:delta P13}.
When we consider the IR limit ($x\ll 1$), the leading contribution of $\delta P^{(22)}$ in the series of $x$ vanishes 
after the integration of $\nu$
and the dominant contribution of $\delta P^{(22)}+\delta P^{(13)}$ is proportional to the linear power spectrum as
\al{
	\delta P^{(22)}({\bm k})+\delta P^{(13)}({\bm k})
		\stackrel{\rm{IR}}{\approx}2Z_1({\bm k})\left(\Upsilon_0 +\Upsilon_{\rm IR}\right)\sigma_{0,{\rm IR}}^2 P_\delta (k)
		\propto k^{n_{\rm IR}}
	\,,
}
where $\sigma_{0,{\rm IR}}^2:=\int_{p\ll k}\dd pp^2P_\delta (p)/2\pi^2$ and $\Upsilon_{\rm IR}$ represents
the scale-independent coefficient.
Since $P_{\rm std}^{(22)}+P_{\rm std}^{(13)}\stackrel{\rm{IR}}{\propto}k^2P_\delta\propto k^{2+n_{\rm IR}}$, 
the sum of the one-loop correction terms at small scales decays faster than the standard terms and does not dominate the signal.
On the other hand, in the case of the opposite limit ($x\gg 1$), the situation changes.
$\delta P^{(22)}$ approaches to constant (see Fig.~\ref{fig:pow_z30_mu0}), but $\delta P^{(13)}$ gives
the nontrivial scale-dependence as 
\al{
	\delta P^{(13)}({\bm k})
		\stackrel{\rm{UV}}{\approx}2Z_1({\bm k})\left(\Upsilon_0 +\Upsilon_{\rm UV}\right)\sigma_{0,{\rm UV}}^2 P_\delta (k)
	\,,\label{eq:shift1}
}
where $\sigma_{0,{\rm UV}}^2:=\int_{p\gg k}\dd pp^2P_\delta (p)/2\pi^2$ 
and $\Upsilon_{\rm UV}$ represents the scale-independent coefficient given by
\al{
	\Upsilon_{\rm UV}
		=-\left(\frac{8}{35}\alpha_1+\frac{72}{245}\overline T_{21}\right)+\left(\frac{58}{105}\alpha_1+\frac{148}{245}\overline T_{21}\right)\mu^2
	\,.
}
This expression shows that $\delta P^{(13)}$ is proportional to the linear power spectrum even at large scales, 
thus it corresponds to the scale-independent shift of the prefactor $Z_1^2({\bm k})$ in the linear term.
\\

Finally, the nontrivial contribution from the primordial bispectrum is expressed as
\al{
	P^{(12)}({\bm k})
		=&2Z_1({\bm k})\frac{k^3}{(2\pi)^2}\int_0^\infty\dd xx^2\int_{-1}^{+1}\dd\nu
			\Bigl[\Xi (x,\nu ;\mu)+\Sigma (x,\nu ;\mu)\Bigr]
	\notag\\
	&\qquad\times
			{\cal M}(k){\cal M}(kx){\cal M}\bigl( k\sqrt{1+x^2-2\nu x}\,\bigr)
			B_\zeta (k,kx,k\sqrt{1+x^2-2\nu x}\,)
	\,,\label{eq:P12 def}
}
where $\Xi$ and $\Sigma$ were defined in Eqs.~\eqref{eq:Xi def} and \eqref{eq:Sigma def}.
At small scales, the nonlinear growth of structure becomes quickly important as mildly nonlinear scales are approached,
and $P^{(12)}$ cannot give the significant contribution to the 21-cm power spectrum.
We then focus only on larger scales, in which the UV contribution $x\gg 1$ becomes dominated.
For scales of our interest, the leading contribution due to the primordial bispectrum is given by
\al{
	P^{(12)}({\bm k})
		\stackrel{\rm{UV}}{\approx} 2Z_1({\bm k})\left(\alpha_2^{(1)}+\frac{1}{3}\alpha_1+\frac{1}{5}\overline T_{21}\right)
				{\cal M}(k)\int_{p\gg k}\frac{\dd p}{2\pi^2}p^2{\cal M}^2(p)B_\zeta (k,p,p)
	\,.\label{eq:P12 UV limit}
}
When we consider the local form of the primordial bispectrum defined in Eq.~\eqref{eq:local fNL def},
Eq.~\eqref{eq:P12 UV limit} can reduce to
\al{
	P^{(12)}({\bm k})
		\stackrel{\rm{UV}}{\approx} \frac{24}{5}Z_1({\bm k})\frac{f_{\rm NL}^{\rm local}}{{\cal M}(k)}
			\left(\alpha_2^{(1)}+\frac{1}{3}\alpha_1+\frac{1}{5}\overline T_{21}\right)\sigma_{0,{\rm UV}}^2 P_\delta (k)
	\,.\label{eq:shift2}
}
This expression contains a term inversely proportional to ${\cal M}(k)\propto k^2$, which leads to the strong scale-dependence at large scales,
as observed in Figs.~\ref{fig:pow_z30_mu0} and \ref{fig:pow}.
An interesting observation from Fig.~\ref{fig:pow} is that the redshift dependence of $P^{(12)}$ at $z=30$--$50$ is weaker than that of $P^{(11)}$, 
presumably because the redshift dependence is estimated by $a^3(z)\alpha_1^2(z)$ and $\alpha_1(z)$ roughly decays as $a^{-1.5}(z)$ during these phase.

In the case of non-local models of primordial non-Gaussianity, 
the bispectrum in the squeezed limit, $B_\zeta (k,p,|{\bm k}-{\bm p}|)$ with $k\ll p$, are 
asymptotically given by $B_\zeta\propto (k/p)P_\zeta (k)P_\zeta (p)$ for the orthogonal type and 
$B_\zeta\propto (k/p)^2P_\zeta (k)P_\zeta (p)$ for the equilateral type (see e.g., \cite{Matsubara:2012nc}).
Hence, on sufficiently large scales, $P^{(12)}$ is proportional to $k/{\cal M}(k)\propto k^{-1}$ and $k^2/{\cal M}(k)\propto k^0$
for the orthogonal and equilateral types, respectively, implying that their scale-dependence
becomes weaker than that of the local model Eq.~\eqref{eq:shift2} and cannot
dominate the signals. 
These features can be used to discriminate between the types of primordial non-Gaussianity.

Combining Eqs.~\eqref{eq:shift1} and \eqref{eq:shift2}, we conclude that the large-scale 21-cm power spectrum including 
the higher-order contributions can be well approximated by the following expression:
\al{
	P({\bm k})
		\stackrel{\rm{UV}}{\approx}\biggl\{
			Z_1^2({\bm k})+2Z_1({\bm k})\biggl[(\text{const.})+\frac{12}{5}\frac{f_{\rm NL}^{\rm local}}{{\cal M}(k)}
			\left(\alpha_2^{(1)}+\frac{1}{3}\alpha_1+\frac{1}{5}\overline T_{21}\right)\biggr]\sigma_{0,{\rm UV}}^2
		\biggr\} P_\delta ({\bm k})
	\,.\label{eq:reduced P}
}
This expression is one of the main results in this paper. 
Since this expression is very similar to the formula of the scale-dependent bias in the context of galaxy surveys, 
we expect that the resultant scale-dependence of the prefactor due to the primordial bispectrum can be used to constrain 
the primordial non-Gaussianity by using observations of the 21-cm power spectrum.
Hereafter, we use Eq.~\eqref{eq:reduced P} as the fiducial model in the subsequent forecast.

\section{Impact on parameter estimation}
\label{sec:Fisher analysis}

\subsection{Fisher-matrix analysis}
\label{sec:Fisher-matrix analysis}

In this section, we apply the Fisher-matrix method to explore the potential impact of the use of the 21-cm power spectrum as well as 
the 21-cm bispectrum to constrain the primordial non-Gaussianity.
Given an antenna array with a baseline $D_{\rm base}$ uniformly covered a fraction $f_{\rm cover}$, observing for a time $t_{\rm obs}$,
the instrumental noise power spectrum can be written as~\cite{Munoz:2016owz,Zaldarriaga:2003du}
\al{
	P_{\rm N}(z)=\frac{\pi T_{\rm sys}^2}{t_{\rm obs}f_{\rm cover}^2}r^2(z)y_\nu (z)\frac{\lambda^2(z)}{D_{\rm base}^2}
	\,,
}
where $\lambda =c(1+z)/\nu_{21}$ is the redshifted wavelength corresponding to 21-cm line,
$r(z)$ is the conformal distance, $y_\nu (z)$ is the conversion function from frequency
to wavenumber parallel to the line-of-sight.
The system temperature $T_{\rm sys}$ is assumed to be dominated by the galactic synchrotron emission, which is parametrized as~\cite{deOliveira-Costa:2008cxd}
\al{
	T_{\rm sys}(\nu )=180\,{\rm K}\times\left(\frac{\nu}{180\,{\rm MHz}}\right)^{-2.6}
	\,.
}
With this noise model, we adopt the Fisher analysis to estimate expected errors of model parameters for a given 21-cm experiment.
We separate the available comoving volume in redshift bins, and then compute the Fisher matrix for one of these slices centered at redshift $z_i$ as
\al{
	F_{\alpha\beta}^{(i)}\approx F_{\alpha\beta}^{{\rm P},(i)}+F_{\alpha\beta}^{{\rm B},(i)}
	\,,
}
where
\al{
	&F_{\alpha\beta}^{{\rm P},(i)}=\sum_{{\bm k}}\frac{\pd P({\bm k};z_i)}{\pd\theta^\alpha}\left[{\rm C}^{-1}_{\rm PP}\right]\frac{\pd P({\bm k};z_i)}{\pd\theta^\beta}
	\,,\\
	&F_{\alpha\beta}^{{\rm B},(i)}=\sum_{{\bm k}_1,{\bm k}_2,{\bm k}_3}
			\frac{\pd B({\bm k}_1,{\bm k}_2,{\bm k}_3;z_i)}{\pd\theta^\alpha}
			\left[{\rm C}^{-1}_{\rm BB}\right]
			\frac{\pd B({\bm k}_1,{\bm k}_2,{\bm k}_3;z_i)}{\pd\theta^\beta}
	\,.
}
Here we neglect the cross covariance between the power- and bi-spectra for simplicity.
In order to take advantage of using the 21-cm line, we would like to add from different redshift slices.
In our tomographic analysis, we take the bandwidth $\Delta\nu =1\,{\rm MHz}$ to keep the information in each redshift slice 
uncorrelated~\cite{Munoz:2015eqa}.
We then approximate the total Fisher matrix by summing over redshifts as
\al{
	F^{\rm tot}_{\alpha\beta}\approx\sum_i F_{\alpha\beta}^{(i)}
	\,.
}
Assuming the Gaussian error covariance, the covariances of the power- and bi-spectra are expected as~\cite{Sefusatti:2007ih}
\al{
	&{\rm C}_{\rm PP}
		=\frac{V_{\rm survey}(z_i)}{N_{\rm P}(z_i)}P_{\rm tot}^2({\bm k};z_i)
	\,,\\
	&{\rm C}_{\rm BB}
		=\frac{s_{\rm B}V_{\rm survey}(z_i)}{N_{\rm B}(z_i)}P_{\rm tot}({\bm k}_1;z_i)P_{\rm tot}({\bm k}_2;z_i)P_{\rm tot}({\bm k}_3;z_i)
	\,.
}
where $P_{\rm tot}$ is the 21-cm power spectrum including the noise contamination given by $P_{\rm tot}({\bm k};z)=P({\bm k};z)+P_{\rm N}(z)$, 
$s_{\rm B}=6,2,1$ for equilateral, isosceles and general triangles, respectively.
The quantities $N_{\rm P}=V_{\rm P}/k_{\rm F}^3$ and $N_{\rm B}=V_{\rm B}/k_{\rm F}^6$ denote the total numbers of available pairs and triangles 
with $k_{\rm F}=2\pi /V_{\rm survey}^{1/3}$ and $V_{\rm P,B}$ being the fundamental frequency and the volume of the fundamental cell in Fourier space.
To take into account the effect of the velocity perturbations, we need to consider the orientation with respect to the line-of-sight, $\mu$.
As for the power spectrum analysis, the volume $V_{\rm P}$ can be taken to be $V_{\rm P}=2\pi k^2\Delta k\Delta\mu$.
On the other hand, the bispectrum becomes a function of five variables; Three of them describe the shape of the triangle
($k_1$, $k_2$, and $k_3$ or the angle $\theta$ between ${\bm k}_1$ and ${\bm k}_2$) and the two remaining variables characterize
the orientation of the triangle with respect to the line-of-sight, $\mu$, and the azimuthal angle $\phi$~\cite{Scoccimarro:1999ed}.
With this parametrization, the three directional cosines are given as
$\mu_1 =\mu$\,, $\mu_2 =\mu\cos\theta-\sqrt{1-\mu^2}\sin\theta\cos\phi$\,, $\mu_3 =-(k_1\mu +k_2\mu_2)/k_3$.
The volume of the fundamental cell for the bispectrum can be written as $V_{\rm B}=2\pi k_1k_2k_3\Delta k_1\Delta k_2\Delta k_3\Delta\mu\Delta\phi$.
In the subsequent analysis, we take $\Delta k=k_{\rm F}$ and $\Delta k_1=\Delta k_2=\Delta k_3=100k_{\rm F}$.

In this paper, we consider several different types of noise level.
As an example for a futuristic radio array (FRA), we assume a baseline of $D_{\rm base}=100\,{\rm km}$, 
a coverage fraction of $f_{\rm cover}=0.2$ or $0.5$, 
a sky coverage of $2\pi$ steradian ($f_{\rm sky}=0.5$), and $10^4$ hours of observations.
We also consider the cosmic-variance limited (CVL) case, in which $P_{\rm N}=0$ to show the theoretical ultimate limits that 
can be observed in this probe. 
The observed maximum perpendicular wavenumber can be determined in terms of the baseline of each array as
\al{
	k_\perp^{\rm max}\approx \frac{2\pi D_{\rm base}}{r(z)\lambda (z)}
	\,.
}
For simplicity, we assume that the radial resolution $k_\parallel^{\rm max}$ matches the angular resolution $k_\perp^{\rm max}$,
while the radial resolution in practice might be easier to achieve through better frequency binning.
The minimum wavenumber is limited by the astrophysical foregrounds.
In this paper we assume that the minimum wavenumber is taken to be~\cite{Mao:2008ug}
\al{
	k_{\rm min}\approx\frac{2\pi}{y_\nu\Delta\nu}
	\,,\label{eq:kMin}
}
where $\Delta\nu$ is the bandwidth for each redshift bin probed by 
the experiment~\footnote{In the several studies, e.g., \cite{Morales:2005qk,Meerburg:2016zdz}, 
the minimum parallel wavenumber is determined by not the bandwidth for each redshift bin 
but the total bandwidth.
Since these studies focused only on the 21-cm bispectrum, the results do not depend sensitively on
the choice of the minimum wavenumber.
However, our results are expected to be sensitive to the value of the minimum wavenumber, because
the contribution from the primordial non-Gaussianity to the 21-cm power spectrum leads to the large-scale enhancement.
Therefore, in this paper we take the conservative choice.
}. 

\begin{figure}
\includegraphics[width=150mm]{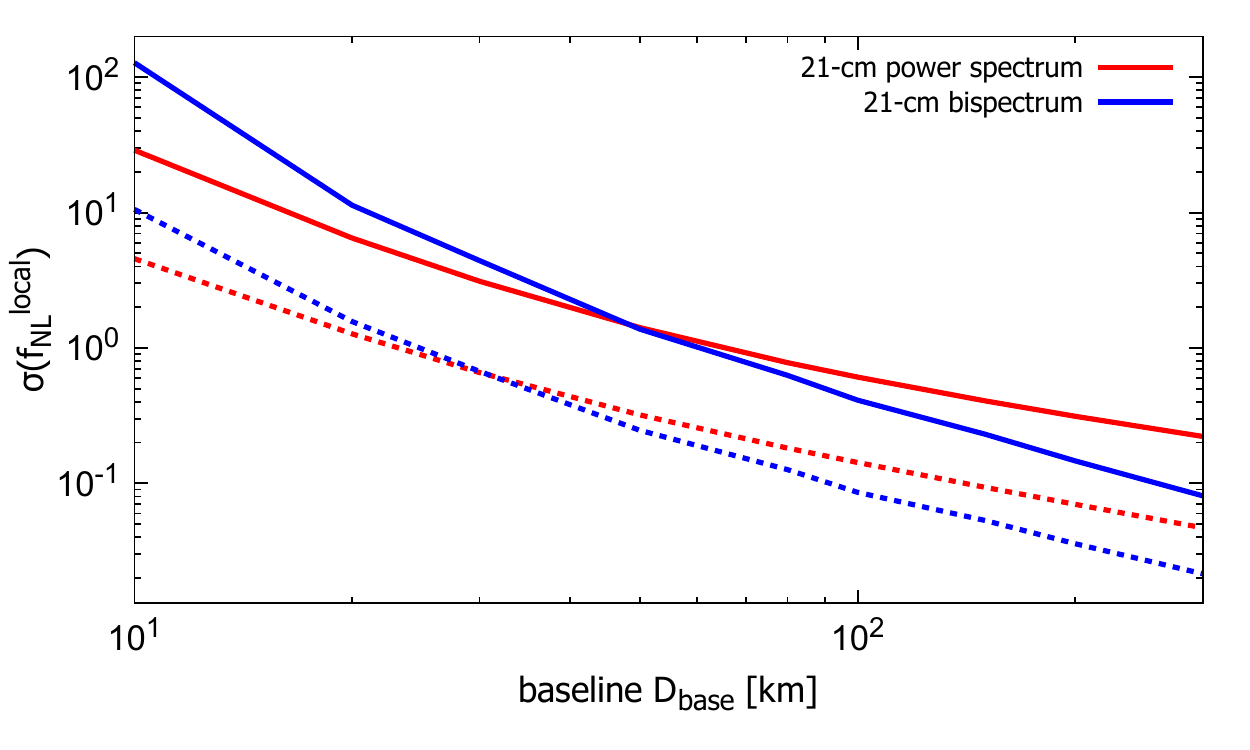}
\vspace{-5mm}
\caption{
The unmarginalized (dashed) and marginalized (solid) $1\sigma$ errors on $f_{\rm NL}^{\rm local}$
as a function of the baseline $D_{\rm base}$.
The different colors represent the different observables: the 21-cm power spectrum (red) and
the 21-cm bispectrum (blue), respectively. 
The other survey parameters are assumed to be 
$f_{\rm cover}=1$\,, $f_{\rm sky}=1$, $t_{\rm obs}=10^4\,{\rm hrs}$, and $\Delta\nu =1\,{\rm MHz}$.
}
\label{fig:sigmafNL}
\end{figure} 

Finally, let us discuss the contribution from the nonlinear growth of structure, which leads to the one-loop correction 
in the 21-cm power spectrum and the secondary bispectrum in the 21-cm bispectrum.
To parametrize the nonlinear evolution, we here model the time-dependence of the coefficients
$\alpha_n^{(m)}(z)$ (we have used the notation $\alpha_0^{(1)}=\overline T_{21}$ and $\alpha_1^{(1)}\equiv\alpha_1$).
Since these are smooth functions of a redshift, these can be modeled by a linear combination of several basis functions.
We parametrize these smooth contributions as a seventh-order polynomial as~\cite{Munoz:2015eqa}
\al{
	\alpha_n^{(m)}(z)=\sum_{j=0}^{7}\alpha_{n,j}^{(m)}[\log (z/50)]^j
	\,.
}
The fitted smooth functions are used as our phenomenological model that captures the dependence of 
the nonlinear contributions.
In what follows, we marginalize over these coefficients in our forecasts.
Therefore, in our Fisher analysis, we consider one parameter for the primordial spectrum $f_{\rm NL}^{\rm local}$, 
the $5\times 8=40$ nuisance parameters $\overline T_{21,j}\,,\alpha_{1,j}\,,\alpha_{2,j}^{(2)}\,,\alpha_{3,j}^{(1)}\,,\alpha_{3,j}^{(2)}$ 
for 21-cm power spectrum,
and the $4\times 8=32$ nuisance parameters $\overline T_{21,j}\,,\alpha_{1,j}\,,\alpha_{2,j}^{(1)}\,,\alpha_{2,j}^{(2)}$ for the 21-cm bispectrum.
On the other hand, we fix standard cosmological parameters to those of standard $\Lambda$CDM model.

\subsection{Results}

Before showing expected constraints from FRA and CVL, let us first estimate the dependence of
the survey parameters. 
In Fig.~\ref{fig:sigmafNL}, we plot the unmarginalized (dashed line) and marginalized (solid line) errors on $f_{\rm NL}^{\rm local}$ 
as a function of the baseline $D_{\rm base}$. 
Different curves represent different observables; the 21-cm power spectrum (red) and the 21-cm bispectrum (blue).
Here, for simplicity we assume the instrumental model $f_{\rm cover}=1$, $f_{\rm sky}=1$, $t_{\rm obs}=10^4\,{\rm hrs}$, and $\Delta\nu =1\,{\rm MHz}$.
This figure implies that the expected constraint from the 21-cm power spectrum 
has the weak dependence on the value of the baseline $D_{\rm base}$, while that from the 21-cm bispectrum is sensitive to $D_{\rm base}$.
This is understood as follows: The increase in the baseline results in both the decrease in the noise and the increase in
the maximum value of the wavenumber. 
Since, as shown in section \ref{sec:One-loop power spectrum of 21-cm fluctuations and signature of primordial non-Gaussianity}, 
the scale-dependence of the 21-cm power spectrum due to the primordial bispectrum can dominate the signals only at very large scales, 
the impact of the increase in the baseline is expected to be relatively small.
On the other hand, as for the 21-cm bispectrum, the effect of $f_{\rm NL}^{\rm local}$ appears even at small scales, hence
the increase in the baseline leads to the increase of the Fourier samples, which can reduce the sample noises.
An interesting observation is that in the case of $D_{\rm base}\lesssim 100\,{\rm km}$ the constraining power of the 21-cm power spectrum
is stronger than that of the 21-cm bispectrum and 
for a reasonable sized array with a baseline of several tens of kilometers the 21-cm power spectrum can reach the important threshold 
$\sigma (f_{\rm NL}^{\rm local})={\cal O}(1)$, while in the ultimate situation, namely $D_{\rm base}\gg 100\,{\rm km}$,
the 21-cm bispectrum becomes more powerful and can reach $\sigma (f_{\rm NL}^{\rm local})\lesssim O(0.1)$, which is consistent with the previous 
results~\cite{Pillepich:2006fj,Munoz:2015eqa,Meerburg:2016zdz,Floss:2022grj}.

\begin{table}[tb]
\caption{Forecast results of marginalized (unmarginalized) $1\sigma$ errors on primordial non-Gaussianity parameters 
$f_{\rm NL}^{\rm local}$.
}
\label{table:forecast}
\vspace{2mm}
\begin{tabular}{|c||c|c|c|} \hline
$\sigma (f_{\rm NL}^{\rm local})$ & Power spectrum & Bispectrum & Combined  \\ \hline\hline
FRA ($f_{\rm cover}=0.2$) & $8.24$ $(1.78)$     & $13.2$ $(1.65)$ & $6.93$ $(1.21)$  \\ \hline
FRA ($f_{\rm cover}=0.5$) & $1.95$ $(0.45)$     & $1.85$ $(0.31)$ & $1.32$ $(0.26)$ \\ \hline
CVL                            & $0.085$ $(0.0078)$ & $0.0084$ $(0.0028)$ & $0.0083$ $(0.0026)$ \\ \hline
\end{tabular}
\end{table}

Finally, Table \ref{table:forecast} shows the expected marginalized (unmarginalized) constraints on $f_{\rm NL}^{\rm local}$
for FRA ($f_{\rm cover}=0.2$ and $0.5$) and CVL. 
We found that the $1\sigma$ errors of the local-type nonlinear parameter in the ideal case are 
$\sigma (f_{\rm NL}^{\rm local})\lesssim{\cal O}(0.1)$ for the 21-cm power spectrum and
$\sigma (f_{\rm NL}^{\rm local})\lesssim{\cal O}(0.01)$ for the 21-cm bispectrum.
On the other hand, in the more realistic situation, the constraining power of the 21-cm power spectrum is of the same order as that of the 21-cm bispectrum
and both the observables can constrain $f_{\rm NL}^{\rm local}$ at the level of $\sigma (f_{\rm NL}^{\rm local})={\cal O}(1)$, 
which is comparable to that from the current CMB measurements.
Therefore, we conclude that the precise measurement of not only the 21-cm bispectrum but also 
the 21-cm power spectrum delivered by very low frequency future radio array can be used to constrain the primordial non-Gaussianity 
and the constraining power of the 21-cm power spectrum is comparable to or severer than that of the 21-cm bispectrum

\section{Summary}
\label{sec:Summary}

In this paper, we have studied the effect of primordial non-Gaussianity and nonlinear growth of structure
on the power spectrum of the 21-cm line differential brightness temperature fluctuation.
Employing the perturbative treatment of gravitational clustering, we have calculated the leading-order
non-Gaussian and one-loop corrections to the 21-cm power spectrum through the nonlinear mode-coupling.
In the weakly nonlinear regime, the leading one-loop term on large scales induces the scale-independent 
shift of the prefactor in the linear power spectrum, while on small scales the standard one-loop terms give
the dominant contribution. 
We have also quantitatively estimated the non-Gaussian signals on the 21-cm power spectrum.
In the local model of primordial non-Gaussianity the non-Gaussian correction
induces the strong scale-dependent enhancement on large scales, while non-local type
primordial non-Gaussianities such as equilateral- and orthogonal-types lead to the weaker
scale-dependence. 
These remarkable properties are useful in constraining local-type primordial non-Gaussianity through 
the 21-cm power spectrum and discriminating between the type of primordial non-Gaussianity.

We then estimate the potential impact of the parameter estimation from future observations
for the 21-cm line measurement, in particular by using the 21-cm power spectrum.
Taking into account both the scale-dependent enhancement due to primordial non-Gaussianity and 
the scale-independent shift due to the one-loop correction in the 21-cm power spectrum, we have performed 
the Fisher-matrix analysis to give the forecasts for constraints of the local-type nonlinear parameter
$f_{\rm NL}^{\rm local}$.
We have shown that for the reasonable size of the baseline of several tens of kilometers
the constraining power of the 21-cm power spectrum is stronger than
that of the 21-cm bispectrum, and the 21-cm power spectrum can reach the threshold value to distinguish 
between single-field and multi-field inflation models, $\sigma (f_{\rm NL}^{\rm local})\lesssim 1$.
In the ideal case, the 21-cm power spectrum can constrain the nonlinear parameter 
at the level of $\sigma (f_{\rm NL}^{\rm local})\lesssim {\cal O}(0.1)$.
Although in the ultimate situation the constraining power of the 21-cm bispectrum eventually becomes significant,
we have shown that the analysis by using the 21-cm power spectrum is expected to be very helpful during earlier stages of observations.

In this paper, we have neglected the effect of the baryon sound speed and have simply used the CDM perturbative kernels 
as the baryon perturbative kernels.
In Ref.~\cite{Floss:2022grj} the authors considered the baryonic pressure when analyzing the 21-cm bispectrum
by using the formula in \cite{Shoji:2009gg}.
Hence, it might also affect the 21-cm power spectrum.
Since our method for deriving the one-loop power spectrum includes the loop integration,
modes with wavelengths close to the baryonic Jeans scale might affect the spectrum due to the mode-coupling.
Moreover, when performing the Fisher matrix analysis, we have taken into account only the large-scale asymptotes of 
the one-loop and non-Gaussian correction terms.
In this sense, our results presented in this paper may be regarded as a conservative constraint and
these are needed to be investigated for a more realistic quantitative estimation.
These prospects are left to be studied in a future work.

\section*{Acknowledgment}

We would like to thank Atsushi Taruya for fruitful discussion.
This work was supported in part by JSPS KAKENHI Grants No.~17K14304, No.~19H01891, and No.~22K03627.

\appendix
\section{Coefficients}
\label{sec:Coefficients}

In this section, we will show the explicit form of the coefficients in Eq.~\eqref{eq:T21 exp}, following Ref.~\cite{Pillepich:2006fj}.
The amplitude of the global signals of the 21-cm line differential brightness temperature is given by
\al{
	\overline T_{21}={\cal T}_0\left( 1-\frac{\overline T_{\rm CMB}}{\overline T_{\rm s}}\right)
	\,,
}
with
\al{
	{\cal T}_0(z)=\frac{3c^3\hbar A_{10}}{16k_{\rm B}\nu_{21}^2}\frac{\overline n_{\rm H}(z)(1-\overline x_{\rm e}(z))}{(1+z)H(z)}
	\,.
}
The coefficients of the 21-cm fluctuations in Eq.~\eqref{eq:T21 exp} are
\al{
	&{\cal T}_{\rm b}
		=\overline T_{21}
			-{\cal T}_0\frac{\overline T_{\rm CMB}(\overline T_{\rm CMB}-\overline T_{\rm gas})\overline y_{\rm c}}{\overline T_{\rm s}^2(1+\overline y_{\rm c})^2}
	\,,\\
	&{\cal T}_{\rm T}
		={\cal T}_0\frac{\overline T_{\rm CMB}\overline y_{\rm c}}{\overline T_{\rm s}(1+\overline y_{\rm c})}
			\biggl[
				1-\frac{\overline T_{\rm CMB}-\overline T_{\rm gas}}{\overline T_{\rm s}(1+\overline y_{\rm c})}\eta_1
			\biggr]
	\,,
}
for the first order,
\al{
	&{\cal T}_{\rm bb}
		=-{\cal T}_0\frac{\overline T_{\rm CMB}^2(\overline T_{\rm CMB}-\overline T_{\rm gas})\overline y_{\rm c}}{\overline T_{\rm s}^3(1+\overline y_{\rm c})^3}
	\,,\\
	&{\cal T}_{\rm TT}
		=-{\cal T}_0\frac{\overline T_{\rm CMB}\overline y_{\rm c}}{\overline T_{\rm s}(1+\overline y_{\rm c})}
			\biggl[
				1-\frac{\overline T_{\rm CMB}}{\overline T_{\rm s}(1+\overline y_{\rm c})}\eta_1
				-\frac{\overline T_{\rm gas}(\overline T_{\rm CMB}-\overline T_{\rm gas})\overline y_{\rm c}}{\overline T_{\rm s}^2(1+\overline y_{\rm c})^2}\eta_1^2
				+\frac{\overline T_{\rm CMB}-\overline T_{\rm gas}}{\overline T_{\rm s}(1+\overline y_{\rm c})}\eta_2
			\biggr]
	\,,\\
	&{\cal T}_{\rm bT}={\cal T}_0\frac{\overline T_{\rm CMB}\overline y_{\rm c}}{\overline T_{\rm s}(1+\overline y_{\rm c})}
			\biggl[
				1+\frac{\overline T_{\rm CMB}}{\overline T_{\rm s}(1+\overline y_{\rm c})}
				-\frac{2\overline T_{\rm CMB}(\overline T_{\rm CMB}-\overline T_{\rm gas})}{\overline T_{\rm s}^2(1+\overline y_{\rm c})^2}\eta_1
			\biggr]
	\,,
}
for the second order, and
\al{
	&{\cal T}_{\rm bbb}
		={\cal T}_0\frac{\overline T_{\rm CMB}^2\overline T_{\rm gas}(\overline T_{\rm CMB}-\overline T_{\rm gas})\overline y_c^2}{\overline T_{\rm s}^4(1+\overline y_{\rm c})^4}
	\,,\\
	&{\cal T}_{\rm TTT}
		={\cal T}_0\frac{\overline T_{\rm CMB}\overline y_{\rm c}}{\overline T_{\rm s}(1+\overline y_{\rm c})}
			\biggl[
				1
				-\frac{\overline T_{\rm CMB}}{\overline T_{\rm s}(1+\overline y_{\rm c})}\eta_1
				-\frac{\overline T_{\rm CMB}\overline T_{\rm gas}\overline y_{\rm c}}{\overline T_{\rm s}^2(1+\overline y_{\rm c})^2}\eta_1^2
				-\frac{(\overline T_{\rm CMB}-\overline T_{\rm gas})\overline T_{\rm gas}^2\overline y_{\rm c}^2}{\overline T_{\rm s}^3(1+\overline y_{\rm c})^3}\eta_1^3
	\notag\\
	&\qquad\qquad\qquad
				+\frac{\overline T_{\rm CMB}}{\overline T_{\rm s}(1+\overline y_{\rm c})}\eta_2
				+\frac{2(\overline T_{\rm CMB}-\overline T_{\rm gas})\overline T_{\rm gas}\overline y_{\rm c}}{\overline T_{\rm s}^2(1+\overline y_{\rm c})^2}\eta_1\eta_2
				-\frac{\overline T_{\rm CMB}-\overline T_{\rm gas}}{\overline T_{\rm s}(1+\overline y_{\rm c})}\eta_3
			\biggr]
	\,,\\
	&{\cal T}_{\rm bbT}
		={\cal T}_0\frac{\overline T_{\rm CMB}^3\overline y_{\rm c}}{\overline T_{\rm s}^3(1+\overline y_{\rm c})^3}
			\biggl[
				1-\frac{(\overline T_{\rm CMB}-\overline T_{\rm gas})(\overline T_{\rm CMB}-2\overline T_{\rm gas}\overline y_{\rm c})}{\overline T_{\rm CMB}\overline T_{\rm s}(1+\overline y_{\rm c})}\eta_1
			\biggr]
	\,,\\
	&{\cal T}_{\rm bTT}
		=-{\cal T}_0
			\frac{2\overline T_{\rm CMB}\overline y_{\rm c}}{\overline T_{\rm s}(1+\overline y_{\rm c})}
			\biggl[
				1-\frac{\overline T_{\rm gas}\overline y_{\rm c}}{2\overline T_{\rm s}(1+\overline y_{\rm c})}
				-\frac{\overline T_{\rm CMB}^2}{\overline T_{\rm s}^2(1+\overline y_{\rm c})^2}\eta_1
	\notag\\
	&\qquad\qquad\qquad
				-\frac{3\overline T_{\rm CMB}\overline T_{\rm gas}(\overline T_{\rm CMB}-\overline T_{\rm gas})\overline y_{\rm c}}{2\overline T_{\rm s}^3(1+\overline y_{\rm c})^3}\eta_1^2
				+\frac{\overline T_{\rm CMB}(\overline T_{\rm CMB}-\overline T_{\rm gas})}{\overline T_{\rm s}^2(1+\overline y_{\rm c})^2}\eta_2
			\biggr]
	\,,
}
for the third order,
where we have neglected the contributions from the perturbations of the hydrogen ionization fraction and the Ly-$\alpha$ pumping efficiency.
Here we have introduced the symbols $\eta_n$ to parametrize the dependence of the collisional coupling as
\al{
	&\kappa_{10}=\overline\kappa_{10}(\overline T_{\rm gas})
			\Bigl\{
				1+\eta_1\delta_{\rm T}+\eta_2[\delta_{\rm T}]^2+\eta_3[\delta_{\rm T}]^3+\cdots
			\Bigr\}
	\,.
}

\section{Evolution equations for matter temperature and ionization fraction}
\label{sec:Evolution equations}

Following Refs.~\cite{Pillepich:2006fj,Munoz:2015eqa,Floss:2022grj}, 
expanding the equation for the gas temperature order-by-order, we obtain the first- and second-order equations given by
\al{
	&\dot\delta_{\rm T}^{(1)}
		=\frac{2}{3}\dot\delta_{\rm b}^{(1)}	
			+\Gamma_{\rm C}\biggl[
				\left(\frac{\overline T_{\rm CMB}}{\overline T_{\rm gas}}-1\right)\delta_x^{(1)}
				-\frac{\overline T_{\rm CMB}}{\overline T_{\rm gas}}\delta_{\rm T}^{(1)}
			\biggr]
	\,,\label{eq:delta_T1 eq}\\
	&\dot\delta_{\rm T}^{(2)}
		=\frac{2}{3}\dot\delta_{\rm b}^{(2)}
			+\Gamma_{\rm C}\biggl[
				\left(\frac{\overline T_{\rm CMB}}{\overline T_{\rm gas}}-1\right)\delta_x^{(2)}
				-\frac{\overline T_{\rm CMB}}{\overline T_{\rm gas}}\delta_{\rm T}^{(2)}
				-\delta_x^{(1)}\delta_{\rm T}^{(1)}
			\biggr]
			+\frac{2}{3}\dot\delta_{\rm b}^{(1)}\left(\delta_{\rm T}^{(1)}-\delta_{\rm b}^{(1)}\right)
	\,,
}
and the third-order equation given by
\al{
	\dot\delta_{\rm T}^{(3)}
		=&\frac{2}{3}\dot\delta_{\rm b}^{(3)}
			+\Gamma_{\rm C}\biggl[
				\left(\frac{\overline T_{\rm CMB}}{\overline T_{\rm gas}}-1\right)\delta_x^{(3)}
				-\frac{\overline T_{\rm CMB}}{\overline T_{\rm gas}}\delta_{\rm T}^{(3)}
				+\delta_x^{(2)}\delta_{\rm T}^{(1)}+\delta_x^{(1)}\delta_{\rm T}^{(2)}
			\biggr]
	\notag\\
	&
			+\frac{2}{3}\dot\delta_{\rm b}^{(2)}\left(\delta_{\rm T}^{(1)}-\delta_{\rm b}^{(1)}\right)
			+\frac{2}{3}\dot\delta_{\rm b}^{(1)}\Bigl[\delta_{\rm T}^{(2)}-\delta_{\rm b}^{(2)}-\delta_{\rm b}^{(1)}\left(\delta_{\rm T}^{(1)}-\delta_{\rm b}^{(1)}\right)\Bigr]
	\,.\label{eq:delta_T3 eq}
}
The equations above shows that $\delta_{\rm T}^{(n)}$ can be solved in the form given in Eqs.~\eqref{eq:delta_X1} and \eqref{eq:delta_X3}.

We expand the evolution equation for the ionization fraction to obtain the equation for $\delta_x^{(n)}$ as
\al{
	&\dot\delta_x^{(1)}
		=-\Gamma_{\rm R}\left(\delta_x^{(1)}+A_1\delta_{\rm T}^{(1)}+\delta_{\rm b}^{(1)}\right)
	\,,\\
	&\dot\delta_x^{(2)}
		=-\Gamma_{\rm R}
			\biggl[
				\delta_x^{(2)}+A_1\delta_{\rm T}^{(2)}+\delta_{\rm b}^{(2)} +[\delta_x^{(1)}]^2+2\delta_x^{(1)}\delta_{\rm b}^{(1)}
				+A_1\delta_{\rm T}^{(1)}\left(\delta_{\rm b}^{(1)}+2\delta_x^{(1)}\right) +A_2[\delta_{\rm T}^{(1)}]^2
			\biggr]
	\,,
}
and
\al{
	\dot\delta_x^{(3)}
		=&-\Gamma_{\rm R}
			\biggl[
				\delta_x^{(3)}+A_1\delta_{\rm T}^{(3)}+\delta_{\rm b}^{(3)}
				+2\delta_x^{(1)}\delta_x^{(2)}+2\left(\delta_x^{(2)}\delta_{\rm b}^{(1)}+\delta_x^{(1)}\delta_{\rm b}^{(2)}\right)+[\delta_{x}^{(1)}]^2\delta_{\rm b}^{(1)}
	\notag\\
	&\qquad\qquad
				+2A_1\left(\delta_x^{(2)}\delta_{\rm T}^{(1)}+\delta_x^{(1)}\delta_{\rm T}^{(2)}\right)+A_1\delta_{\rm T}^{(1)}\delta_x^{(1)}\left(\delta_x^{(1)}+2\delta_{\rm b}^{(1)}\right)
				+A_1\left(\delta_{\rm T}^{(2)}\delta_{\rm b}^{(1)}+\delta_{\rm T}^{(1)}\delta_{\rm b}^{(2)}\right)
	\notag\\
	&\qquad\qquad
				+A_2[\delta_{\rm T}^{(1)}]\left(\delta_{\rm b}^{(1)}+2\delta_x^{(1)}\right) +2A_2\delta_{\rm T}^{(1)}\delta_{\rm T}^{(2)}+A_3[\delta_{\rm T}^{(1)}]^3
			\biggr]
	\,,
}
where $\Gamma_{\rm R}:=\overline\alpha_{\rm B}\overline n_{\rm H}\overline x_e$ is the background recombination rate and
we have used the symbols $\eta_n$ and $A_n$ to parametrize the dependence of the recombination coefficient as
\al{
	&\alpha_{\rm B}=\overline\alpha_{\rm B}(\overline T_{\rm gas})\Bigl\{ 1+A_1\delta_{\rm T}+A_2[\delta_{\rm T}]^2+A_3[\delta_{\rm T}]^3+\cdots\Bigr\}
	\,.
}

Substituting these into Eqs.~\eqref{eq:delta_T1 eq}--\eqref{eq:delta_T3 eq}, these can reduce to the evolution equation for
the coefficients $C_{{\rm T},n}^{(m)}$ as
\al{
	&\dot C_{{\rm T},1}=H\left(\frac{2}{3}-C_{{\rm T},1}\right)
				+\Gamma_{\rm C}\biggl[
				\left(\frac{\overline T_{\rm CMB}}{\overline T_{\rm gas}}-1\right) C_{x,1}
				-\frac{\overline T_{\rm CMB}}{\overline T_{\rm gas}}C_{{\rm T},1}
			\biggr]
	\,,\\
	&\dot C_{{\rm T},2}^{(2)}=2H\left(\frac{2}{3}-C_{{\rm T},2}^{(2)}\right)
				+\Gamma_{\rm C}\biggl[
				\left(\frac{\overline T_{\rm CMB}}{\overline T_{\rm gas}}-1\right) C_{x,2}^{(2)}
				-\frac{\overline T_{\rm CMB}}{\overline T_{\rm gas}}C_{{\rm T},2}^{(2)}
			\biggr]
	\,,\\
	&\dot C_{{\rm T},2}^{(1)}=\frac{2}{3}H\biggl[-1+C_{{\rm T},1}-3C_{{\rm T},2}^{(1)}\biggr]
				+\Gamma_{\rm C}\biggl[
				\left(\frac{\overline T_{\rm CMB}}{\overline T_{\rm gas}}-1\right) C_{x,2}^{(1)}
				-\frac{\overline T_{\rm CMB}}{\overline T_{\rm gas}}C_{{\rm T},2}^{(1)}
				-C_{x,1}C_{{\rm T},1}
			\biggr]
	\,,
}
and
\al{
	&\dot C_{{\rm T},3}^{(3)}=2H\left(\frac{2}{3}-C_{{\rm T},3}^{(3)}\right)
				+\Gamma_{\rm C}\biggl[
				\left(\frac{\overline T_{\rm CMB}}{\overline T_{\rm gas}}-1\right) C_{x,3}^{(3)}
				-\frac{\overline T_{\rm CMB}}{\overline T_{\rm gas}}C_{{\rm T},2}^{(3)}
			\biggr]
	\,,\\
	&\dot C_{{\rm T},3}^{(2)}=\frac{2}{3}H\biggl[ -3+2C_{{\rm T},1}+C_{{\rm T},2}^{(2)}-3C_{{\rm T},3}^{(2)}\biggr]
	\notag\\
	&\qquad\qquad
				+\Gamma_{\rm C}\biggl[
				\left(\frac{\overline T_{\rm CMB}}{\overline T_{\rm gas}}-1\right) C_{x,3}^{(2)}
				-\frac{\overline T_{\rm CMB}}{\overline T_{\rm gas}}C_{{\rm T},3}^{(2)}
				+C_{x,1}C_{{\rm T},2}^{(2)}+C_{x,2}^{(2)}C_{{\rm T},1}
			\biggr]
	\,,\\
	&\dot C_{{\rm T},3}^{(1)}=\frac{2}{3}H\biggl[ 1-C_{{\rm T},1}+C_{{\rm T},2}^{(1)}-3C_{{\rm T},3}^{(2)}\biggr]
	\notag\\
	&\qquad\qquad
				+\Gamma_{\rm C}\biggl[
				\left(\frac{\overline T_{\rm CMB}}{\overline T_{\rm gas}}-1\right) C_{x,3}^{(1)}
				-\frac{\overline T_{\rm CMB}}{\overline T_{\rm gas}}C_{{\rm T},3}^{(1)}
				+C_{x,1}C_{{\rm T},2}^{(1)}+C_{x,2}^{(1)}C_{{\rm T},1}
			\biggr]
	\,.
}
For the ionization fraction, the evolution equations for the coefficients $C_{x,n}^{(m)}$ are given by
\al{
	&\dot C_{x,1}
		=-HC_{x,1}-\Gamma_{\rm R}\left( C_{x,1}+A_1C_{{\rm T},1}+1\right)
	\,,\\
	&\dot C_{x,2}^{(2)}
		=-2HC_{x,2}^{(2)}-\Gamma_{\rm R}\left( C_{x,2}^{(2)}+A_1C_{{\rm T},2}^{(2)}+1\right)
	\,,\\
	&\dot C_{x,2}^{(1)}
		=-2HC_{x,2}^{(1)}
			-\Gamma_{\rm R}\biggl[
				C_{x,2}^{(1)}+A_1C_{{\rm T},2}^{(1)}
				+C_{x,1}\left( C_{x,1}+2\right)+A_1C_{{\rm T},1}\left(1+2C_{x,1}\right)+A_2[C_{{\rm T},1}]^2
			\biggr]
	\,,
}
and
\al{
	&\dot C_{x,3}^{(3)}
		=-3HC_{x,3}^{(3)}-\Gamma_{\rm R}\left( C_{x,3}^{(3)}+A_1C_{{\rm T},3}^{(3)}+1\right)
	\,,\\
	&\dot C_{x,3}^{(2)}
		=-3HC_{x,3}^{(2)}
			-\Gamma_{\rm R}\biggl[
				C_{x,3}^{(2)}+A_1C_{{\rm T},3}^{(2)}
				+2C_{x,2}^{(2)}\left( 2+C_{x,1}\right)+2C_{x,1}
	\notag\\
	&\qquad\qquad\qquad
				+A_1C_{{\rm T},1}\left( 1+2C_{x,2}^{(2)}\right)
				+A_1C_{{\rm T},2}^{(2)}\left( 1+2C_{x,1}\right)
				+2A_2C_{{\rm T},1}C_{{\rm T},2}^{(2)}
			\biggr]
	\,,\\
	&\dot C_{x,3}^{(1)}
		=-3HC_{x,3}^{(2)}
			-\Gamma_{\rm R}\biggl[
				C_{x,3}^{(1)}+A_1C_{{\rm T},3}^{(1)}
				+2C_{x,2}^{(1)}\left( 1+C_{x,1}\right) +[C_{x,1}]^2
	\notag\\
	&\qquad\qquad\qquad
				+A_1C_{{\rm T},2}^{(1)}\left( 1+2C_{x,1}\right)
				+A_1C_{{\rm T},1}\left( [C_{x,1}]^2+2C_{x,2}^{(1)}+2C_{x,1}\right)		
	\notag\\
	&\qquad\qquad\qquad
				+A_2[C_{{\rm T},1}]^2\left( 1+2C_{x,1}\right)
				+2A_2C_{{\rm T},1}C_{{\rm T},2}^{(1)}
				+A_3[C_{x,1}]^3
			\biggr]
	\,.
}

\section{One-loop power matter spectrum}
\label{sec:Standard one-loop power matter spectrum}

In this section, we briefly summarize the perturbative kernels and present the explicit expression for the one-loop power spectrum,
following Ref.\cite{Bernardeau:2001qr}.
The symmetrized second- and third-order kernels are given by
\al{
	&F_2^{({\rm s})}({\bm k}_1,{\bm k}_2)=\frac{5}{7}+\frac{1}{2}(\widehat{\bm k}_1\cdot\widehat{\bm k}_2)\left(\frac{k_1}{k_2}+\frac{k_2}{k_1}\right)
			+\frac{2}{7}(\widehat{\bm k}_1\cdot\widehat{\bm k}_2)^2
	\,,\label{eq:F2}\\
	&G_2^{({\rm s})}({\bm k}_1,{\bm k}_2)=\frac{3}{7}+\frac{1}{2}(\widehat{\bm k}_1\cdot\widehat{\bm k}_2)\left(\frac{k_1}{k_2}+\frac{k_2}{k_1}\right)
			+\frac{4}{7}(\widehat{\bm k}_1\cdot\widehat{\bm k}_2)^2
	\,,\label{eq:G2}
}
and
\al{
	&F_3^{({\rm s})}({\bm k}_1,{\bm k}_2,{\bm k}_3)
		=\frac{7}{54}\frac{{\bm k}\cdot{\bm k}_1}{k_1^2}F_2^{({\rm s})}({\bm k}_2,{\bm k}_3)
			+\frac{1}{54}\left( 2k^2\frac{{\bm k}_1\cdot{\bm k}_{23}}{k_1^2k_{23}^2}
			+7\frac{{\bm k}\cdot{\bm k}_{23}}{k_{23}^2}\right) G_2^{({\rm s})}({\bm k}_2,{\bm k}_3)+(\text{perms})
	\,,\label{eq:F3}\\
	&G_3^{({\rm s})}({\bm k}_1,{\bm k}_2,{\bm k}_3)
		=\frac{1}{18}\frac{{\bm k}\cdot{\bm k}_1}{k_1^2}F_2^{({\rm s})}({\bm k}_2,{\bm k}_3)
			+\frac{1}{18}\left( 2k^2\frac{{\bm k}_1\cdot{\bm k}_{23}}{k_1^2k_{23}^2}
			+\frac{{\bm k}\cdot{\bm k}_{23}}{k_{23}^2}\right) G_2^{({\rm s})}({\bm k}_2,{\bm k}_3)+(\text{perms})
	\,.\label{eq:G3}
}
With these, the standard contributions from the nonlinear kernels to the 21-cm power spectrum are given by 
\al{
	&P^{(22)}_{\rm std}({\bm k})
		=[\alpha_2^{(2)}]^2P^{(22)}_{\delta\delta}(k)
			-2\alpha_2^{(2)}\overline T_{21}\mu^2 P^{(22)}_{\delta v}(k)
			+[\overline T_{21}]^2\mu^4P^{(22)}_{vv}(k)
	\,,\\
	&P^{(13)}_{\rm std}({\bm k})
		=Z_1({\bm k})\left( \alpha_3^{(3)}P^{(13)}_{\delta\delta} (k)-\overline T_{21}\mu^2P^{(13)}_{vv}(k)\right)
	\,.
}
The explicit expressions of each components are given by (e.g., \cite{Makino:1991rp})
\al{
	&P^{(22)}_{\delta\delta}(k)
		:=2\frac{k^3}{(2\pi)^2}\int_0^\infty\dd x x^2P_\delta (kx)\int_{-1}^1\dd\nu 
			P_\delta \bigl(k\sqrt{1+x^2-2\nu x}\,\bigr)
			\biggl[\frac{3x+7\nu-10\nu^2x}{14x(1+x^2-2\nu x)}\biggr]^2
	\,,\\
	&P^{(13)}_{\delta\delta} (k)
		:=\frac{k^3}{(2\pi)^2}P_\delta (k)\int_0^\infty\dd xx^2P_\delta (kx)
	\notag\\
	&\qquad\qquad\qquad\times
			\frac{1}{252x^2}\biggl[
				\frac{12}{x^2}-158+100x^2-42x^4+\frac{3}{x^3}(x^2-1)^3(7x^2+2)\ln\biggl|\frac{x+1}{x-1}\biggl|\,
			\biggr]
	\,.
}
for the density field,
\al{
	&P^{(22)}_{vv}(k)
		:=2\frac{k^3}{(2\pi)^2}\int_0^\infty\dd x x^2P_\delta (kx)\int_{-1}^1\dd\nu 
			P_\delta \bigl(k\sqrt{1+x^2-2\nu x}\,\bigr)
			\biggl[\frac{-x+7\nu -6\nu^2 x}{14x(1+x^2-2x\nu )}\biggr]^2
	\,,\\
	&P^{(13)}_{vv}(k)
		:=\frac{k^3}{(2\pi)^2}P_\delta (k)\int_0^\infty\dd xx^2P_\delta (kx)
	\notag\\
	&\qquad\qquad\qquad\times
			\frac{1}{84x^2}\biggl[
				\frac{12}{x^2}-82+4x^2-6x^4+\frac{3}{x^3}(x^2-1)^3(x^2+2)\ln\biggl|\frac{x+1}{x-1}\biggl|\,
			\biggr]
	\,,
}
for the velocity divergence field, and
\al{	
	P^{(22)}_{\delta v}(k)
		:=&2\frac{k^3}{(2\pi)^2}\int_0^\infty\dd x x^2P_\delta (kx)\int_{-1}^1\dd\nu 
			P_\delta \bigl(k\sqrt{1+x^2-2\nu x}\,\bigr)
	\notag\\
	&\qquad\qquad\times
			\biggl[\frac{3x+7\nu-10\nu^2x}{14x(1+x^2-2\nu x)}\frac{-x+7\nu -6\nu^2 x}{14x(1+x^2-2x\nu )}\biggr]
	\,,
}
for the cross term of the density and velocity divergence field.
\\

Moreover, to show the explicit expression of the terms including the orientation with respect to the line-of-sight, 
it is convenient to use
\al{
	&{\bm k}=(0,0,k)
	\,,\ \ \ 
	\widehat{\bm n}=(0,\sqrt{1-\mu^2},\mu )
	\,,\\
	&{\bm p}=p(\sqrt{1-\nu^2}\cos\phi ,\sqrt{1-\nu^2}\sin\phi ,\nu )
	\,.
}
The coordinates $\nu$ and $\phi$ were defined in Sec.~\ref{sec:Fisher-matrix analysis}.
With these notations, the following integration with respect to $\phi$ can be calculated as
\al{
	&\int_0^{2\pi}\frac{\dd\phi}{2\pi}\mu_{\bm p}^2
		=\frac{1}{2}(1-\mu^2)(1-\nu^2)+\mu^2\nu^2
	\,,\label{eq:integral1}\\
	&\int_0^{2\pi}\frac{\dd\phi}{2\pi}\mu_{{\bm k}-{\bm p}}^2
		=\frac{x^2(1-\mu^2)(1-\nu^2)+2\mu^2(1-x\nu)^2}{2(1+x^2-2x\nu)}
	\,,\\
	&\int_0^{2\pi}\frac{\dd\phi}{2\pi}\mu_{\bm p}^4
		=\frac{3}{8}(1-\mu^2)^2(1-\nu^2)^2+3\mu^2\nu^2(1-\mu^2)(1-\nu^2)+\mu^4\nu^4
	\,,\label{eq:integral3}
}
and
\al{
	&\int_0^{2\pi}\frac{\dd\phi}{2\pi}\mu_{\bm p}^2\mu_{{\bm k}-{\bm p}}^2
		=\frac{3x^2(1-\mu^2)^2(1-\nu^2)^2+4\mu^2(1-\mu^2)(1-\nu^2)(1-6x\nu +6x^2\nu^2)+8\mu^4\nu^2(1-x\nu)^2}{8(1+x^2-2x\nu)}
	\,.
}
Based on these results, 
we write down the explicit expression of the last term in \eqref{eq:delta P13} as
\al{
	\delta P^{(13)}({\bm k})
		\supset -Z_1({\bm k})\frac{k^3}{(2\pi)^2}P_\delta (k)\int_0^\infty\dd x x^2P_\delta (kx)
				\Bigl[\alpha_1{\cal G}_1(x;\mu)+2\overline T_{21}\mu^2{\cal G}_2(x;\mu)\Bigr]
	\,.\label{eq:G2 contribution}
}
By using Eqs.~\eqref{eq:integral1}--\eqref{eq:integral3}, we obtain the functional form of ${\cal G}_1$ and ${\cal G}_2$ as
\al{
	&{\cal G}_1(x;\mu )=6\int_{-1}^{+1}\dd\nu\int_0^{2\pi}\frac{\dd\phi}{2\pi}\,
		\frac{1}{3}\Bigl[\mu_{{\bm k}+{\bm p}}^2G_2({\bm k},{\bm p})+\mu_{{\bm k}-{\bm p}}^2G_2({\bm k},-{\bm p})\Bigr]
	\notag\\
	&\qquad
		=\frac{1}{168}
			\biggl[
			 -\frac{18(1-3\mu^2)}{x^2}+66+634\mu^2-6x^2\left( 3x^2-11\right)\left( 1-3\mu^2\right)
	\notag\\
	&\qquad\qquad
			+\frac{9}{x^3}\left( x^2-1\right)^4\left( 1-3\mu^2\right)\ln\biggl|\frac{x+1}{x-1}\biggl|\,
			\biggr]
	\,,\\
	&{\cal G}_2(x;\mu)=6\int_{-1}^{+1}\dd\nu\int_0^{2\pi}\frac{\dd\phi}{2\pi}\,
		\frac{1}{3}\mu_{\bm p}^2\Bigl[\mu_{{\bm k}+{\bm p}}^2G_2({\bm k},{\bm p})+\mu_{{\bm k}-{\bm p}}^2G_2({\bm k},-{\bm p})\Bigr]
	\notag\\
	&\qquad
		=\frac{1}{4480x^5}
			\biggl[
				6x\left(1+x^2\right)\left( 15-100x^2+298x^4-100x^6+15x^8\right)
	\notag\\
	&\qquad
				+4x\left( 15+35x^2+2334x^4-1410x^6+915x^8-225x^{10}\right)\mu^2
	\notag\\
	&\qquad
				+10x\left( x^2-1\right)^2\left( 9+15x^2-145x^4+105x^6\right)\mu^4
	\notag\\
	&\qquad
				+15\left( x^2-1\right)^4
					\Bigl\{
						3\left(x^2-1\right)^2+2\left(1+6x^2-15x^4\right)\mu^2
						+\left( 3+10x^3+35x^4\right)\mu^4
					\Bigr\}\ln\biggl|\frac{x+1}{x-1}\biggl|\,
			\biggr]
	\,.
}

\end{document}